\documentclass[amstex,amsfonts,aps,prb,overcite,amsmath]{revtex4}
\usepackage{graphicx}

\begin{document}
\title{\textbf{Treatment of Linear and Nonlinear Dielectric Property of Molecular Monolayer and Submonolayer
with Microscopic Dipole Lattice Model: I. Second Harmonic Generation
and Sum-Frequency Generation}}
\author{\textsf{De-sheng Zheng}\footnote[2]{Also Graduate School of the Chinese Academy of Sciences.}}
\author{\textsf{Yuan Wang\footnote[3]{An exchange graduate student at
the Institute of Chemistry, the Chinese Academy of Sciences from the
Department of Physics, Anhui Normal University, Wuhu, China
241000.}}}
\author{\textsf{An-an Liu\footnotemark[2]}}
\author{\textsf{Hong-fei Wang}\footnote[1]{Author to whom
correspondence should be addressed. E-mail: hongfei@iccas.ac.cn.
Tel: 86-10-62555347. Fax: 86-10-62563167.}}\affiliation{State Key
Laboratory of Molecular Reaction Dynamics,
\\Institute of Chemistry, the Chinese Academy of Sciences,
Beijing, China 100080}

\date{\today}
\begin{abstract}

Two crucial issues in the interface studies with the interface
specific nonlinear spectroscopy are on how to describe the
macroscopic dielectric constant of the molecularly thin layer and
how to calculate the anisotropic two dimensional microscopic local
field effect. In the currently accepted models of the nonlinear
optics, the nonlinear radiation was treated as the result of an
infinitesimally thin polarization sheet layer, and a three layer
model was generally employed. The direct consequence of this
approach is that an apriori dielectric constant, which still does
not have a clear definition, has to be assigned to this polarization
layer. Because the Second Harmonic Generation (SHG) and the
Sum-Frequency Generation vibrational Spectroscopy (SFG-VS) have been
proven as the sensitive probes for interfaces with the submonolayer
coverage, the treatment based on the more realistic discrete induced
dipole model needs to be developed. Here we show that following the
molecular optics theory approach the SHG, as well as the SFG-VS,
radiation from the monolayer or submonolayer at an interface can be
rigorously treated as the radiation from an induced dipole lattice
at the interface. In this approach, the introduction of the
polarization sheet is no longer necessary. Therefore, the ambiguity
of the unaccounted dielectric constant of the polarization layer is
no longer an issue. Moreover, the anisotropic two dimensional
microscopic local field factors can be explicitly expressed with the
linear polarizability tensors of the interfacial molecules. Based on
the planewise dipole sum rule in the molecular monolayer, crucial
experimental tests of this microscopic treatment with SHG and SFG-VS
are discussed. Many puzzles in the literature of surface SHG and SFG
spectroscopy studies can also be understood or resolved in this
framework. This new treatment may provide a solid basis for the
quantitative analysis in the surface SHG and SFG studies.
\end{abstract}
\pacs{42.65.Ky, 68.18.Jk}

\maketitle

\section{Introduction}

The past two and half decades have witnessed tremendous advancement
and applications of the interface specific second order nonlinear
optical techniques, mainly the surface second harmonic generation
(SHG) and sum-frequency generation vibrational spectroscopy
(SFG-VS), in molecular interface
studies.\cite{ShenARMS1986,ShenNature1989,Richmond1988Review,ShenARPC1989,ShenVogeARMS1991,
EisenthaARPC1992,EisenthalACR1993,Corn1994ChemRev,cdbain:jcsf1995,Eisenthal1996ChemRev,
Shen1999JPCB103p3292,SomorjaiJPCB1999,ShultzIRPC2000,ShenIEEE2000,RichmondARPC2001,
BuckJVSTA2001,ShenPAC2001,RichmondCR2002,SomorjaiChenARPC2002,TadjeddineRPP2005,WangHFIRPC2005,
ShenChemRev2006,AllenChemRev2006,EisenthalChemRev2006,WangHFPCCP2006}
To simply put it, SHG is the second order nonlinear process where
two photons with the same fundamental frequency ($\omega$) interact
with a nonlinear medium simultaneously to generate a photon with the
second harmonic frequency ($2\omega$). If the two fundamental
frequencies are not the same, a photon at the sum of these two
frequencies can be generated from the so-called SFG process. Because
of the symmetry requirement for the second order nonlinear
processes, the leading dipolar term of the SHG or SFG processes is
generally forbidden for the centrosymmetric bulk medium, and thus
SHG and SFG become effective probes for the interface between the
two centrosymmetric bulk phases.\cite{ShenNonlinearOpticsBook} The
theoretical foundation and the experimental demonstration of the
interfacial selectivity of SHG as well as SFG were pioneered by Shen
and his colleagues since the early
1980's,\cite{ShenYR1981PRL46p145,ShenYRHeinzTF1982PRL48p478,
Shen1983PRA28p1883,Shen1985JVSTBRichmond40,GuyotsionnestPRB19868254,
ShenZhuPRB1987,Shen1988CPL,ShenFellerPRA1991} extending from the
original formulation of the light waves at the boundary of nonlinear
media by Bloembergen \textit{et al.} in the early
1960's.\cite{BloembergenPR1962606}

With SHG and SFG-VS, equilibrium and dynamic behaviors of the
molecular interface or film can be directly measured from the
nonlinear electronic or vibrational spectroscopic response of the
interfacial molecular
moieties.\cite{ShenNature1989,EisenthaARPC1992,Corn1994ChemRev,
Eisenthal1996ChemRev,Shen1999JPCB103p3292,EisenthalChemRev2006} One
particular aspect of the studies in the past few years has been
focused on the quantitative measurement and interpretation of the
molecular orientation and vibrational spectra from the SHG and
SFG-VS measurement on various molecular
interfaces.\cite{SimpsonRowlwnACR2000,WangRaoJCP2003,WangHFIRPC2005,WangHFPCCP2006}
Recent works also demonstrated that the coherent nature of the
surface SHG and SFG-VS processes makes them more advantages over the
other incoherent spectroscopic techniques used in surface studies.
The interference of the molecular electric field and the strong
polarization dependence in the SHG and SFG-VS response can be
employed to investigate the detailed interactions and to determine
the molecular conformation at the molecular
interfaces.\cite{Ganweiwater2,GanweiJPCC2007p8716,GanweiJPCC2007p8726}

However, researchers have long realized that the crucial issues in
the quantitative interface studies with the SHG and SFG are on how
to describe the macroscopic dielectric constant of the molecularly
thin layer and how to calculate the anisotropic two dimensional
microscopic local field
effect.\cite{HaydenPRB19883718,CnossenLangmuir19931974,MunnJCP199310059,
ZhangJOSAB1990902,CnossenJCP19924512,McGilpSyntheticMetal1993181,
TangJPCM19933791,UiJCP19946430,EisertPRB199810860,ZhuangPRB199912632,WeiXing2000PRE62p5160,
MishinaJCP20024016,RowlenAnalChem20025954,WangHFIRPC2005,WangHFPCCP2006}
It has been well demonstrated that in majority cases the calculation
of the molecular orientation is quite sensitive to the values used
for these two factors.\cite{ZhangJOSAB1990902,EisertPRB199810860,
RowlenAnalChem20025954,WangHFIRPC2005,WangHFPCCP2006}

In the currently accepted models of the nonlinear optics of
interface,\cite{Heinzthesis1982,HeinzBook,ShenARPC1989,ShenFellerPRA1991,
ZhuangPRB199912632,WeiXing2000PRE62p5160} the nonlinear radiation
was treated as the result of an infinitesimally thin polarization
sheet layer, and as the starting pint a macroscopic dielectric
constant was assigned to this thin
layer.\cite{Heinzthesis1982,HeinzBook,ShenARPC1989,ShenFellerPRA1991}
This model was a natural extension of the original formulation by
Bloembergen and Pershan, where a nonlinear plane parallel slab with
a finite thickness embedded between two linear dielectrics was the
source of the SHG radiation, and where the Maxwell's equations which
satisfy the boundary conditions at the two plane interfaces were
accurately solved.\cite{BloembergenPR1962606} However, according to
that treatment, Bloembergen and Pershan concluded that the surface
dipolar contribution to the SHG signal should be overwhelmed by the
quadrupolar radiation in the much thicker boundary layer. Later
experimental observations demonstrated that the surface dipolar
contribution could be dominant in many cases. By solving the
Maxwell's equations which satisfy the boundary conditions at the
infinitesimally thin polarization sheet layer, Heinz and Shen
provided the theoretical basis for the interface specific SHG, as
well as the SFG-VS, in the interface
studies.\cite{Heinzthesis1982,HeinzBook,ShenARPC1989,ShenFellerPRA1991}

Parallel to this macroscopic treatment, Ye and Shen employed a
classical induced point-dipole model to include the microscopic
local field effect on the nonlinear optical properties of adsorbed
molecules on a substrate.\cite{YePRB19834288} Shen later realized
that from the theoretical point of view, the dielectric constant in
the macroscopic model is not well defined for a monolayer because it
is a macroscopic or mesoscopic property. Thus, Shen \textit{et al.}
tried to phenomenologically interpret this macroscopic dielectric
constant as a result of the microscopic local field correction in a
monolayer, and they also gave the explicit expressions for the
macroscopic Fresnel factors and the microscopic local-field factors
.\cite{ZhuangPRB199912632,WeiXing2000PRE62p5160} Zhuang \textit{et
al.} also demonstrated that in a Langmuir monolayer it is
satisfactory to treat the whole molecule with one unique microscopic
local field factor derived from a modified Lorentz model of the
interface.\cite{ZhuangPRB199912632} This approach has been widely
used for interpretation of the SHG and the SFG-VS data
since.\cite{WangHFIRPC2005,WangHFPCCP2006}

However, puzzles and confusions remained in the SHG and SFG-VS
practices on how the linear macroscopic dielectric constant and the
microscopic local field factors, as described by  Ye and Shen,
should be used in the quantitative treatment of the experimental
data. In their SHG measurement of the self-assembled-monolayer (SAM)
at the gold substrate, Eisert \textit{et al.} carefully compared the
results with the macroscopic three layer model and the two layer
model, as well as the local field corrections. They concluded that
using the two-layer model without local-field correction gave most
satisfactory agreement of the molecular orientation with the results
from the NEXAFS and IR spectroscopy
measurements.\cite{EisertPRB199810860} This work is obviously not
consistent with the above treatment by Shen \textit{et al.} It was
further criticized by Roy on the inconsistency that the SAM film is
considered anisotropic for SHG but assumed to be isotropic in terms
of its linear optical properties.\cite{RoyPRB200013283} Roy then
outlined a macroscopic phenomenological model that treats linear and
nonlinear optics of the anisotropic interface layer in a single
framework, which nevertheless requires a minimum of six unknown
parameters of the interfacial optical constants even for an uniaxial
monolayer in SHG or
SFG-VS.\cite{RoyPRB200013283,SimpsonAnalChem2005215} Thus, Roy
concluded that the conventional SHG measurement of phase and
intensity may not be enough to determine the orientation of SAM.
However, even though the anisotropy is undoubtedly significant on
the nonlinear responses, Roy's criticism might have exaggerated the
influence of the anisotropy on the linear optical property within
the monolayer, which in fact only slightly modifies the reflectivity
of the surface as judged from ellipsometry
measurement.\cite{EllipsometryBook,HeavensThinFilmBook,GuyotsionnestApllPhyB1987237,ShenNonlinearOpticsBook}

There have been a few experimental and empirical theory studies on
the SHG data from the interfacial monolayer, using different
parameters of the macroscopic three layer model plus the microscopic
local field
factors.\cite{HaydenPRB19883718,CnossenLangmuir19931974,MunnJCP199310059,
CnossenJCP19924512,McGilpSyntheticMetal1993181,
TangJPCM19933791,UiJCP19946430,EisertPRB199810860,MishinaJCP20024016,
BloembergenBooK196569,LupoJOSAB1988300,HsiungCPL1989539} In dealing
with the macroscopic dielectric constant, generally one of the two
bulk phase values were used, reducing the macroscopic problem to the
same as the so-called two phase
model.\cite{SipeJOSAB1987,SipeJOSAB1988} However, because the two
phase model does not consider the different microscopic properties
of the surface molecular layer, it can only phenomenologically
describe the surface SHG process, and it generally fails to
qualitatively explain the observed differences in the SHG
measurement. On the other hand, in dealing with the microscopic
local field factors, some works followed the classical dipole model
as presented by Ye and Shen, and the rest followed the simple
Lorentz-Lorenz local field expression using the bulk
polarizabilities.\cite{LupoJOSAB1988300,HsiungCPL1989539}

Another approach completely neglects the treatment of the
microscopic local field effect, and treated the molecular film with
a macroscopic three layer
model.\cite{RowlenAnalChem20025954,ZhangJOSAB1990902,SimpsonAnalChem2005215,
ChenJCPB2004,FreyCPL2000454,NaujokJPC1996,HigginsCornLangmuir1992,BainCPL1995}
In SFG-VS, some considered the film anisotropic and its optical
constant is determined using the Clausius-Mossotti relationship with
the ellipsometry data of the film.\cite{BainCPL1995} This is
obviously rooted in the practices of the ellipsometry studies, where
it is generally believed that the macroscopic optical model can be
valid down to the monolayer
level.\cite{EllipsometryBook,HeavensThinFilmBook} In SHG, some
considered the film isotropic and the optical constant can be
obtained from the Kramers-Kronig dispersion relationship measured
from the bulk or thin film UV-Visible absorption
spectra.\cite{NaujokJPC1996,HigginsCornLangmuir1992,ZhangJOSAB1990902}
When the molecule is considered with uniaxial symmetry, the ratio
between the dielectric constants of the fundamental and the SH
frequency can be directly determined from the intensity ratio in the
polarization measurements.\cite{ZhangJOSAB1990902,WangRaoJCP2003}
There are also efforts trying to show that the dielectric constant
of the extremely thin monolayer is just the simple arithmetic
average of the two bulk dielectric
constants.\cite{RowlenAnalChem20025954,SimpsonAnalChem2005215,
ChenJCPB2004,FreyCPL2000454} Event though the results from these
approaches seemed reasonable, neglecting the anisotropic microscopic
local field effect in the molecular monolayer is not likely to be
physically sound.

All the above approaches gave the similar trend in the variation of
molecular orientation. However, the value of the orientational angle
depends quite significantly on the different parameters used.
Besides all the different approaches in the SHG and SFG-VS data
analysis, the experimental studies have shown that the initial
treatment by Heinz and Shen with the infinitesimally thin
polarization sheet layer is generally correct. Researchers in the
field also agree that the definition of the dielectric constant of
the monolayer or even submonolayer film is unclear and meaningless
in the macroscopic optics, because when the film is as thin as one
molecular layer the macroscopic reflection and transmission
coefficients simply converge to the bare interface
limit.\cite{EllipsometryBook,HeavensThinFilmBook,SimpsonAnalChem2005215}
Therefore, even though the treatment provided by Shen and coworkers
is generally valid, there are still issues need to be clarified. One
problem in the existing model is that the macroscopic and
microscopic linear optical properties of the nonlinear molecular
layer are not totally compatible. In practice we also need a more
detailed microscopic theory to help understand the detailed SHG or
SFG-VS responses from different molecular interface or film.

Aside from the above, a microscopic theory of molecular crystal
surface SHG was proposed by Munn in the early
1990's.\cite{MunnJCP1995} In this formulation, since the surface SHG
response is treated microscopically as a sum of responses from
successive surface layers, the introduction of a dielectric constant
of the surface layer was not required, and this incidentally yielded
a microscopic expression of the dielectric constant of the surface
layer. Subsequent studies by Munn and coworkers used the planewise
sum rules in the molecular crystal dielectric
theory\cite{SchacherPR1965,MahanPR1969,PhilpottJCP1972588,PhilpottJCP1972996,PhilpottJCP1973595,PhilpottJCP19745306}
to simulate the linear and nonlinear optical responses of the model
Langmuir-Blodgett
films.\cite{MunnJCP199310052,MunnJCP199310059,PanhuisJCP20006763,PanhuisJCP200010685,PanhuisJCP200010691}
In these treatment, Munn \textit{et al.} concluded that improper
treatment of the local fields can result in significant errors in
determination of the molecular angle from the SHG
data.\cite{PanhuisJCP20006763,PanhuisJCP200010691} They also showed
that in order not to overestimate the microscopic local field
factors for the closely packed monolayer films, the monolayer itself
had to be segmented into several
layers.\cite{MunnJCP199310052,MunnJCP199310059} In the microscopic
theory, the microscopic local field factors depend on the
orientation and distribution of the interfacial dipoles, and the
microscopic local field factors are needed to determine the dipole
orientation from the SHG data. Therefore, Pannhuis and Munn
suggested that a self-consistent approach needs to be employed to
solve the problem.\cite{PanhuisJCP200010691} These microscopic
theory and simulations certainly provided new insights on the
treatment of the SHG from the molecular monolayer as well as
multilayers. These insights may contain answers to the questions
raised above. However, the implications of these works are yet to be
picked up by the field.

Because the SHG and the SFG-VS have been proven as the sensitive
probes for interfaces with the submonolayer
coverage,\cite{ShenNature1989,ShenARPC1989,Corn1994ChemRev,Eisenthal1996ChemRev}
the treatment based on the more realistic discrete induced dipole
model needs to be developed. According to Born and Wolf, the
molecular optics theory can directly connect the macroscopic optical
phenomena to the molecular properties, and can provide deeper
physical insight into electromagnetic interaction problems than does
the rather formal approach based on Maxwell's phenomenological
equations.\cite{Bookmaxborn1997,WolfJOSA1972} From the above
analysis, we realized that different from the infinitesimally thin
polarization sheet layer model treated on the basis of the Maxwell
theory, a molecular optics treatment can be developed to describe
the coherent SHG or SFG-VS radiation from a discrete lattice in
between the two isotropic phases. We shall show in this report that
even though there is no general methods to solve the
integro-differential equations of molecular optics, the summation,
or integration, of the radiation in the far field from a discrete
induced dipole lattice can be rigorously solved with the application
of the principle of the stationary phase.\cite{Bookmaxborn1997}

The discrete induced dipole lattice model is a more realistic
description of the molecular interface than the infinitesimally thin
polarization sheet layer model.\cite{GhinerPRA1994} Because the SHG
and the SFG-VS processes are known with interface selectivity, the
whole problem of the integro-differential equations is greatly
simplified since the summation over the second harmonic or
sum-frequency radiation fields from the discrete interface induced
dipoles needs to be calculated. According to the Ewald-Oseen
extinction theorem in the molecular optics, there exists only the
incident field in vacuum and the dipolar microscopic radiation field
in vacuum emitting from each induced
dipole.\cite{Bookmaxborn1997,WolfJOSA1972,BloembergenPR1962606}
Since the whole problem is treated microscopically as radiation from
the discrete induced dipoles, and the total radiation is calculated
through summation, or integration, over the whole dipole lattice,
there is no need to use the Maxwell equations together with the
boundary conditions as in the infinitesimally thin polarization
sheet layer model. This works because as Lalor and Wolf pointed out
that the Ewald-Oseen extinction theorem just plays the role of the
boundary condition.\cite{WolfJOSA1972} Therefore, there is no need
to use the boundary conditions, where a macroscopic dielectric
constant has to be defined in order to apply the boundary conditions
for a finite volume across the boundary area which includes the
molecular monolayer. Therefore, the macroscopic dielectric constant
for the molecular monolayer, which has difficulties to be defined in
the previous treatment, is just a nuisance parameter in this
microscopic, and it is finally able to be gotten rid of. There was a
section in their classic paper where Bloembergen and
Pershan,\cite{BloembergenPR1962606} showed that the approach using
the integral equation based on the Ewald-Oseen extinction theorem
reached exactly the same results for the nonlinear response from the
nonlinear plate parallel
slab.\cite{BloembergenPR1962606,BloembergenPhysRev19621918} This
certainly helps to justify the microscopic molecular optics approach
to be employed.

In the next few sections, we shall show that the SHG radiation in
the far field from the discrete dipole lattice is in the same form
as from the infinitesimally thin polarization sheet layer model.
Moreover, in this discrete induced dipole model, the introduction of
the polarization sheet is no longer necessary, therefore, the
ambiguity of the unaccounted dielectric constant of the polarization
layer is no longer an issue. Incidentally, the anisotropic two
dimensional microscopic local field factors can be explicitly
expressed with the linear polarizability tensors of the interfacial
molecules. Based on the planewise dipole sum rule in the molecular
monolayer, crucial experimental tests of this treatment with the SHG
and SFG-VS experiments are going to be discussed. We shall also
discuss the puzzles in the literature of the surface SHG and SFG
spectroscopy studies as discussed above. This microscopic treatment
can provide a solid basis for the quantitative analysis of the
surface SHG and SFG studies.

\section{The induced dipole of the monolayer with the discrete point-dipole lattice
model}\label{sectionII}

Because the SHG and the SFG-VS have been proven as the sensitive
probes for interfaces with the submonolayer
coverage,\cite{ShenNature1989,ShenARPC1989,Corn1994ChemRev,Eisenthal1996ChemRev}
the treatment based on the more realistic discrete induced dipole
model needs to be developed. Here we derive the expressions of the
linear and nonlinear induced dipole of the monolayer with the
discrete point-dipole lattice model. Some aspects of the discussions
below can be found in the earlier paper by Ye and
Shen.\cite{YePRB19834288}

In the classical molecular optics, the response of the medium to the
incident field is described by means of the electric-dipole moments
that are induced in the molecules of the medium under the action of
the incident field.\cite{Bookmaxborn1997,WolfJOSA1972} When an
optical field at a frequency $\omega$ is incident on a medium, it
creates induced dipole at the incident frequency and its higher
harmonics in that medium through the total field that each dipole or
molecule experiences. This total field is called the local field
$\vec{E}_{loc}$. The higher harmonic induced dipole can be viewed as
the results of multiple interactions with the local optical field.
Therefore, they are only strong enough for detection when the
optical field is intense enough. The induced dipole at the incident
frequency is the source of the radiation in the linear processes,
while the others are the source responsible for the radiation in the
nonlinear processes. Simply to put it, one has

\begin{eqnarray}
 \vec{\mu}_{induced}  = \vec{\mu}^{linear}_{induced} + \vec{\mu}^{non-linear}_{induced}
\end{eqnarray}

Nonlinear optics has been an extensively studied field since the
invention of the first laser in the early
1960's.\cite{BloembergenIEEE2000,ShenNonlinearOpticsBook} Here we
only discuss the linear and the second harmonic process in an
optical medium. Considering the fact that both fields in the linear
frequency and the resulted second harmonic can contribute to the
induced dipole at the second harmonic frequency, one has

\begin{eqnarray}
 \mu _i  &=& \mu _i^\omega   + \mu _i^{2\omega } \nonumber\\
  &=& (\begin{array}{*{20}c}
   {\alpha _{ix}^\omega } & {\alpha _{iy}^\omega } & {\alpha _{iz}^\omega  }  \\
\end{array})\left( {\begin{array}{*{20}c}
   {E_{loc,x}^\omega  }  \\
   {E_{loc,y}^\omega  }  \\
   {E_{loc,z}^\omega  }  \\
\end{array}} \right)\nonumber\\
&& + ~(\begin{array}{*{20}c}
   {E_{loc,x}^\omega  } & {E_{loc,y}^\omega  } & {E_{loc,z}^\omega  }  \\
\end{array})\left( {\begin{array}{*{20}c}
   {\beta _{ixx} } & {\beta _{ixy} } & {\beta _{ixz} }  \\
   {\beta _{iyx} } & {\beta _{iyy} } & {\beta _{iyz} }  \\
   {\beta _{izx} } & {\beta _{izy} } & {\beta _{izz} }  \\
\end{array}} \right)\left( {\begin{array}{*{20}c}
   {E_{loc,x}^\omega  }  \\
   {E_{loc,y}^\omega  }  \\
   {E_{loc,z}^\omega  }
\end{array}} \right)\nonumber\\
&& + (\begin{array}{*{20}c}
   {\alpha _{ix}^{2\omega} } & {\alpha _{iy}^{2\omega} } & {\alpha _{iz}^{2\omega} }  \\
\end{array})\left( {\begin{array}{*{20}c}
   {E_{loc,x}^{2\omega } }  \\
   {E_{loc,y}^{2\omega } }  \\
   {E_{loc,z}^{2\omega } }  \\
\end{array}} \right)\nonumber\\
\label{dipolemomentatanyfrequency}
\end{eqnarray}

Here $\alpha _{ij}$ is the linear polarizability tensor and $\beta
_{ijk}$ is the second order nonlinear polarizability tensor of the
each molecule in the laboratory coordinates $\lambda(x,y,z)$ with
the index $ijk$ as either one of the three laboratory Cartesian
coordinates. The first term in the
Eq.\ref{dipolemomentatanyfrequency} is the linear induced dipole,
the second term is the induced dipole at the second harmonic
frequency ($2\omega$)induced by the field with the fundamental
frequency $\omega$ through the second harmonic process, and the
third term is the induced dipole at $2\omega$ induced by the field
with $2\omega$ through a linear process. The field with $2\omega$ in
the third term is the result of the second term. The molecular
polarizability tensors in the laboratory coordination
$\lambda(x,y,z)$ can be projected from the molecular polarizability
tensors in the molecular coordination $\lambda'(a,b,c)$.

The above is general for any dielectric medium. In the problem we
describe below, only the induced dipoles at the interface with the
fundamental and second harmonic frequency are considered. The case
for the sum-frequency works similarly following the same approach.
For the isotropic molecular monolayer, there are two independent
linear optical polarizability tensors in the laboratory coordination
$\lambda(x,y,z)$, $\alpha_{xx}= \alpha_{yy}$, and $\alpha_{zz}$; and
there are three non-vanishing independent $\beta$ elements,
$\beta_{xxz}=\beta_{yyz}=\beta_{xzx}=\beta_{yzy}$,
$\beta_{zxx}=\beta_{zyy}$ and
$\beta_{zzz}$.\cite{ZhuangPRB199912632,WangHFPCCP2006} Now the
induce dipole moment of each molecule can be written explicitly as

\begin{eqnarray}
 \mu _x^\omega   &=& \alpha _{xx}^\omega  E_{loc,x}^\omega  \nonumber \\
 \mu _y^\omega   &=& \alpha _{yy}^\omega  E_{loc,y}^\omega  \nonumber \\
 \mu _z^\omega   &=& \alpha _{zz}^\omega  E_{loc,z}^\omega  \nonumber \\
 \mu _x^{2\omega }  &=& \alpha _{xx}^{2\omega } E_{loc,x}^{2\omega }  + \beta _{xxz} E_{loc,x}^\omega  E_{loc,z}^\omega   + \beta _{xzx} E_{loc,z}^\omega  E_{loc,x}^\omega   \nonumber\\
 \mu _y^{2\omega }  &=& \alpha _{yy}^{2\omega } E_{loc,y}^{2\omega }  + \beta _{yyz} E_{loc,y}^\omega  E_{loc,z}^\omega   + \beta _{yzy} E_{loc,z}^\omega  E_{loc,y}^\omega   \nonumber\\
 \mu _z^{2\omega }  &=& \alpha _{zz}^{2\omega } E_{loc,z}^{2\omega }  + \beta _{zxx} E_{loc,x}^\omega  E_{loc,x}^\omega   + \beta _{zyy} E_{loc,y}^\omega  E_{loc,y}^\omega   + \beta _{zzz} E_{loc,z}^\omega
 E_{loc,z}^\omega \label{polarization tensor}
\end{eqnarray}

Now, in order to calculate these induced dipoles of the monolayer,
the knowledge of the local fields $\vec{E}^{\omega}_{loc}$ and
$\vec{E}^{2\omega}_{loc}$ at the interface needs to be described.

When an optical field is incident on a molecular monolayer at the
interface between two bulk phases with dielectric constant
$\varepsilon_{1}=n^{2}_{1}$ and $\varepsilon_{2}=n^{2}_{2}$, the
induced dipole in the monolayer and the image induced dipole in the
substrate all add to the local field at each individual molecule
within the monolayer. Therefore the total local field is different
from the applied electric field.
Then,\cite{BagchiPRL19801475,BagchiPRB19827086,YePRB19834288}

\begin{equation}
\vec E_{loc} (\vec r) = \vec E(\vec r) + \vec E_{dip} (\vec r) +
\vec E_{dip,I} (\vec r)\label{Localfeil}
\end{equation}

Here, let $\vec r$ be the position vector within the monolayer from
the center of the molecule at the origin position. Then $\vec
E_{loc} (\vec r)$ is the local field at $\vec r$, and it is the sum
of three contributions. Here $\vec E(\vec r)$ is the applied field,
$\vec E_{dip} (\vec r)$ is the field acted at this point from all
the induced dipoles except the one at $\vec r$, and $\vec E_{dip,I}
(\vec r)$ is the field created by all the image induced dipoles at
$\vec r$.

Here we first discuss the first term $\vec E(\vec r)$. In this case,
a field of plane wave at frequency $\omega$ is incident at an angle
of $\alpha_{i}$ upon the interface from one bulk phase and reflected
from the interface. Because the applied field is not with frequency
$2\omega$, one always have $\vec E(\vec r,2\omega)=0$. For $\vec
E(\vec r, \omega)$, because of the existence of the interface, the
applied electric field is the superposition of the incident field
plus the field reflected from the
interface.\cite{BagchiPRL19801475,BagchiPRB19827086,YePRB19834288}
According to the Fresnel formulae,\cite{Bookmaxborn1997} the total
applied field $\vec E(\vec r, \omega)$ ($E_{x}$,$E_{y}$,$E_{z}$) can
be determined from the incident electric field
$\vec{E}_{0}$($E_{0,x}$,$E_{0,y}$,$E_{0,z}$)=$E_{0}\hat{e}$ as
followings:
\begin{eqnarray}
 E_x &=& \left( {1 - r_{p}e^{i2k_{1}d\cos \Omega_{i} } } \right)E_{x,0}  = \left( {1 + \frac{{n_1 \cos \Omega_t  - n_2 \cos \Omega_i }}{{n_2 \cos \Omega_i  + n_1 \cos \Omega_t }}e^{i2k_{1}d\cos \Omega_{i} } } \right)E_{x,0}  = L'_{xx} E_{x,0}  \nonumber \\
 E_y &=& \left( {1 + r_{s}e^{i2k_{1}d\cos \Omega_{i} } } \right)E_{x,0}  = \left( {1 + \frac{{n_1 \cos \Omega_i  - n_2 \cos \Omega_t }}{{n_1 \cos \Omega_i  + n_2 \cos \Omega_t }}e^{i2k_{1}d\cos \Omega_{i} } } \right)E_{y,0}  = L'_{yy} E_{y,0}  \nonumber \\
 E_z &=& \left( {1 + r_{p}e^{i2k_{1}d\cos \Omega_{i} } } \right)E_{x,0}  = \left( {1 + \frac{{n_2 \cos \Omega_i  - n_1 \cos \Omega_t }}{{n_2 \cos \Omega _i  + n_1 \cos \Omega_t }}e^{i2k_{1}d\cos \Omega_{i} } } \right)E_{z,0}  = L'_{zz} E_{z,0}  \label{MacroLocalfeil}
\end{eqnarray}

This may be simplified as $\vec{E}=[\mathbb{L}'\cdot \hat{e}]E_{0}$.
Here, $\Omega_{i}$ and $\Omega_{t}$ are the incident angle and the
refraction angle at the interface. $k_{1}$ is the wave vector in
first bulk phase, $d$ is the distance of the dipole from the
interface. Generally for the monolayer $d\ll \lambda$,
$e^{i2k_{1}d\cos \Omega_{i}}\sim 1$. In this case, the expression of
the Fresnel factor $L'_{ii}$ is simplified into $L_{ii}$, which will
be expressed in the Section \ref{SectionIII}.

The expression of $\vec E(\vec r)$ in the Eq.\ref{MacroLocalfeil} is
independent from the structure of the interfacial monolayer for any
interface between two isotropic bulk phases. However, the
expressions of the other two terms in the Eq.\ref{Localfeil} depend
on the model of the monolayer structure. In order to evaluate them,
here a square lattice model is used to represent the molecular
monolayer. Other kinds of lattice models, such as the hexagonal
lattice model, can also be employed and they shall generate similar
results.\cite{ToppingJPRSLA1927,IwamotoPRB19968186} Here we only
discuss the case using the square lattice model.

\begin{center}
\begin{figure}
\includegraphics[width=7cm]{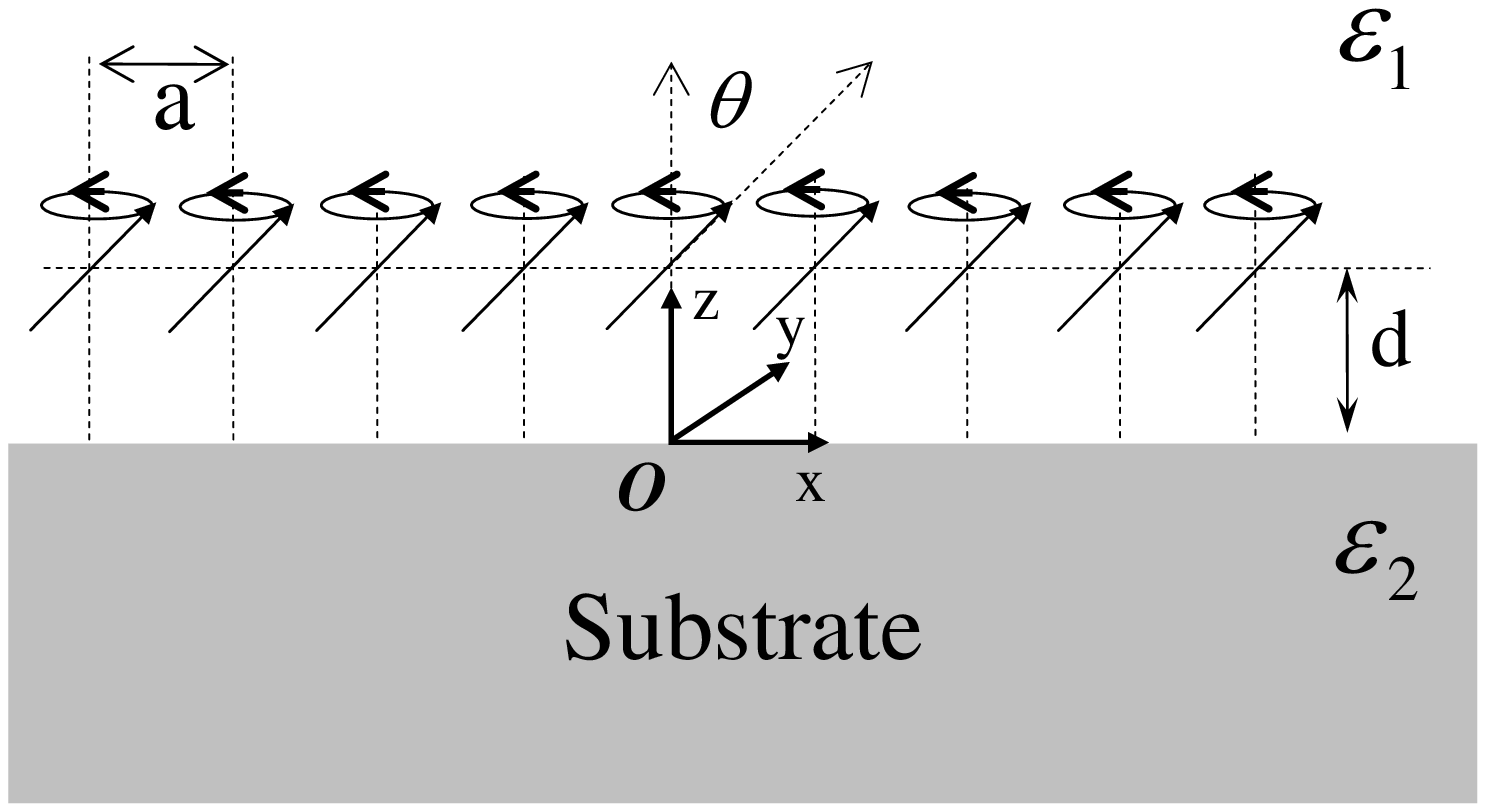}
\caption{Illustration of an infinite lattice of point induced
dipoles on a substrate interface. $a$ is the distance between the
nearest molecules, or the lattice constant; $d$ is the distance from
the center of the point-dipole to the surface of the substrate;
$\theta$ is the tilt angle of the
point-dipoles.}\label{film-point-dipole-model1}
\end{figure}
\end{center}

Bagchi et al \cite{BagchiPRL19801475,BagchiPRB19827086} and Ye et al
\cite{YePRB19834288} used the classical discrete square point-dipole
model to discuss the linear and nonlinear optical responses of the
molecular monolayer, respectively. Such a square lattice model is
illustrated in Fig.\ref{film-point-dipole-model1}. The dipoles at
the interface forms a two-dimensional square lattice with a lattice
constant $a$. The two bulk phases occupy the two semi-infinite space
of $z\geq 0$ and $z \leq 0$ respectively, and they are characterized
by the bulk dielectric constant $\epsilon_{1}$ and $\epsilon_{2}$
(the substrate), respectively. The tilt angle of the dipole with the
surface normal is $\theta$ and its azimuthal orientation is randomly
distributed in the plane of the interface. The distance of the point
dipole to the lower substrate surface is $d$. The coordinates of
each point dipole is then $\vec{R}_{m,n}=(ma,na,d)$, where $m$ and
$n$ are integer indexes along the $x$ and $y$ directions from the
origin $O(0,0,0)$ in the plane of the monolayer, respectively.

The contribution to the dipolar field $\vec E_{dip} (\vec r)$ at an
adsorbed dipole comes from the neighboring induced dipoles within
the monolayer.\cite{BagchiPRB19827086} As in previous
treatments,\cite{BagchiPRL19801475,BagchiPRB19827086,YePRB19834288}
the near field approximation of the local field is employed. In the
near field approximation, the magnetic part of the electromagnetic
field is ignored, and the retardation effects are discarded.
\cite{PhilpottJCP1973595,BagchiPRL19801475,BagchiPRB19827086} This
assumption is justified because the wavelength $\lambda$ of the
incident electric field is much larger than the lattice constant
$a$. Following the works by Bagchi, the electric field at the
lattice point $R_{00}(0,0,d)$ which is created by the rest of the
induced-dipoles in the lattice plane
is\cite{BagchiPRL19801475,BagchiPRB19827086}

\begin{eqnarray}
 E_{dip}^{R_{00} }  &=& \sum\limits_{m,n = -\infty}^{\infty} {'\ \ \frac{{3\left[ {\vec \mu  \cdot (\vec R_{mn}  - \vec R_{00} )} \right](\vec R_{mn}  - \vec R_{00} ) - \vec \mu (\vec R_{mn}  - \vec R_{00} )^2 }}{{(\vec R_{mn}  - \vec R_{00} )^5 }}} \nonumber \\
                    &=& \hat z(\frac{{\mu _z }}{{a^3 }})\xi _0  - \left[ {\hat x(\frac{{\mu _x }}{{a^3 }}) + \hat y(\frac{{\mu _y }}{{a^3 }})} \right]\frac{{\xi _0 }}{2} \label{electric filed of dipole}
\end{eqnarray}

The CGS unit system is used in this paper, as in many textbooks.
Here $\mu_{i}$ is the components of the induced dipole $\vec{\mu}$.
$\hat{x}$, $\hat{y}$, and $\hat{z}$ are the unit vectors in the
laboratory coordinates. The symbol prime $'$ denotes that the
summation does not include the point of origin $R_{00}$. The
constant $\xi _0$ was evaluated by Topping in
1927.\cite{ToppingJPRSLA1927}
\begin{equation}
\xi _0  =  - \sum\limits_{m,n =  - \infty }^\infty  {'\ \ (m^2  +
n^2 )^{ - \frac{3}{2}} }  =  - 9.0336
\end{equation}

In the Eq.\ref{electric filed of dipole}, only the induced dipole in
the same plane of the interface lattice is considered. To be more
rigorous, molecular layers other than the interface lattice plane
also need to be considered. These layers may be the molecules or
atoms of the two bulk phases, whose polarizabilities differ from
those of the dipoles in the interfacial layers. Phenomenologically,
let's define a vertical lattice index $l$ and lattice constant $d$,
with $l=0$ for the plane of the monolayer rather than the plane with
the origin $O(0,0,0)$. Then the total $\vec E_{dip} (\vec
r)=E_{dip}^{R_{00}}+E_{dip}^{R_{00},l\neq 0}$, with
$E_{dip}^{R_{00},l\neq 0}$ as the induced dipole interaction
contribution to the $R_{00}^{0}=R_{00}$ origin point from the
induced dipoles other than the interface lattice plane.

\begin{eqnarray}
 E_{dip}^{R_{00},l\neq 0}  &=& \sum_{l=-\infty}^{\infty}{'}\sum\limits_{m,n = -\infty}^{\infty}{\frac{{3\left[ {\vec \mu^{l}  \cdot (\vec R_{mn}^{l}  - \vec R_{00}^{0} )} \right](\vec R_{mn}^{l}  - \vec R_{00}^{0} ) - \vec \mu^{l} (\vec R_{mn}^{0}  - \vec R_{00}^{0} )^2 }}{{(\vec R_{mn}^{l}  - \vec R_{00}^{0} )^5 }}} \nonumber \\
                    &=& \sum_{l=-\infty}^{\infty}{'\ \ }\bigg( \hat z(\frac{{\mu_{z}^{l} }}{{a_{l}^{3} }})\xi _{0}^{l}  - \left[ {\hat x(\frac{{\mu_{x}^{l} }}{{a_{l}^{3} }}) + \hat y(\frac{{\mu_{y}^{l} }}{{a_{l}^{3} }})} \right]\frac{{\xi _{0}^{l} }}{2}\bigg) \label{OtherElectricFiledofDipole}
\end{eqnarray}

Here $a_{l}$ is the lattice constant of the $l$ layer. The
expression here is similar to the expressions for the image dipole
term by Bagchi and Ye et
al.\cite{BagchiPRB19827086,BagchiPRL19801475,YePRB19834288} Here
$\xi _{0}^{l}$ can be directly derived from the Eq.\ref{electric
filed of dipole}.

\begin{equation}
\xi _0^l  = \sum\limits_{m,n =  - \infty }^\infty  {\frac{{3\left(
{2ld/a_{l}} \right)^2  - \left[ {m^2  + n^2  + \left( {2ld/a_{l}}
\right)^2 } \right]}}{{\left[ {m^2  + n^2  + \left( {2ld/a_{l}}
\right)^2 } \right]^{5/2} }}}\label{dipolesumoftheirlayer}
\end{equation}

To evaluate how quickly this summation converges with increase of
$l$, it can be converted into a more rapidly convergent
series.\cite{BagchiPRB19827086,BagchiPRL19801475}

\begin{equation}
\xi _{0}^{l}  = 16\pi ^2 \sum\limits_{m = 0}^\infty  {\sum\limits_{n
= 1}^\infty  {\left( {m^2  + n^2 } \right)^{1/2} \exp \left[  -
{4\pi l \left( \frac{d}{a_{l}} \right)\left( {m^2  + n^2 }
\right)^{1/2} } \right]} }\label{dipolesumoftheirlayerconvergent}
\end{equation}

The Eq.\ref{dipolesumoftheirlayerconvergent} indicates that $\xi
_{0}^{l}$ falls off exponentially as the $l$ increases. Generally,
when $d \sim a/2$, even the contribution of the immediate
neighboring layer, i.e. $l=\pm 1$ can be neglected. This is the
basis for the planewise dipole sum rule in the molecular crystal
theory, which shall be discussed in the Section
\ref{SectionIV}.\cite{SchacherPR1965,MahanPR1969,PhilpottJCP1972588,
PhilpottJCP1972996,PhilpottJCP1973595,PhilpottJCP19745306} The
planewise sum rule in the molecular crystal theory concluded that
the contribution of the induced dipoles from neighboring layers
contribute insignificantly to the local field of the induced dipoles
in the layer under consideration, comparing to that from the induced
dipoles in the same layer.

The image dipole contribution is just a special case in the
Eq.\ref{OtherElectricFiledofDipole}. Bagchi and Ye \textit{et al.}
discussed and evaluated the image dipole contributions
previously.\cite{BagchiPRL19801475,BagchiPRB19827086,YePRB19834288}
The image dipoles is located in the substrate at
$\vec{R}'_{mn}=(ma,nb,-d)$, i.e. $l=-2$ in the general case above,
and the image dipole is defined
as\cite{BagchiPRL19801475,BagchiPRB19827086,YePRB19834288}
\begin{eqnarray}
 \mu_{I}(\omega)_{x}  &=& \frac{\epsilon_{2} -\epsilon_{1}}{\epsilon_{2} +\epsilon_{1}}(-\mu_{x})\nonumber \\
 \mu_{I}(\omega)_{y}  &=& \frac{\epsilon_{2} -\epsilon_{1}}{\epsilon_{2} +\epsilon_{1}}(-\mu_{y})\nonumber \\
 \mu_{I}(\omega)_{z}  &=& \frac{\epsilon_{2} -\epsilon_{1}}{\epsilon_{2} +\epsilon_{1}}(\mu_{z})
\end{eqnarray}

Thus, the image dipole contribution can also be calculated as the
same procedure as in the
Eq.\ref{OtherElectricFiledofDipole}.\cite{BagchiPRL19801475,BagchiPRB19827086,YePRB19834288}
Ye and Shen concluded that for the typical case on a metal surface,
where $(\epsilon_{2} -\epsilon_{1})/(\epsilon_{2} +\epsilon_{1})
\sim 1$, when $d\geq 2.5{\AA}$ with $a=5{\AA}$, the image induced
dipole contribution is negligible.\cite{YePRB19834288} Therefore, it
is generally accepted that for the dielectric substrate, where
$(\epsilon_{2} -\epsilon_{1})/(\epsilon_{2} +\epsilon_{1}) \ll 1$,
this image induced dipole term needs not to be
considered.\cite{HaydenPRB19883718,McGilpSyntheticMetal1993181,
CnossenLangmuir19931974,TangJPCM19933791,CnossenJCP19924512,UiJCP19946430,
MunnJCP199310059,MishinaJCP20024016,PanhuisJCP200010685,EisertPRB199810860}
Therefore, we shall neglect the image induced dipole term in the
followed treatment.

It has to be noted that both the $\vec E_{dip} (\vec r)$ and the
$\vec E_{dip,I} (\vec r)$ terms have the same expressions for the
fundamental and second harmonic frequencies.

With all the three terms known from above, put the
Eq.\ref{Localfeil}, the Eq.\ref{electric filed of dipole} the
Eq.\ref{OtherElectricFiledofDipole} into the Eq.\ref{polarization
tensor}, one has
\begin{eqnarray}
 \mu _x^\omega   &=& \alpha _{xx}^\omega \l_{xx}(\omega)E_{x}  \nonumber \\
 \mu _y^\omega   &=& \alpha _{yy}^\omega \l_{yy}(\omega)E_{y}  \nonumber \\
 \mu _z^\omega   &=& \alpha _{zz}^\omega \l_{zz}(\omega)E_{z}  \nonumber \\
 \mu _x^{2\omega }  &=& \l_{xx}(2\omega)\beta _{xxz} \l_{xx}(\omega) \l_{zz}(\omega)E_{x}^{\omega} E_{z}^{\omega}    + \l_{xx}(2\omega)\beta _{xzx} \l_{zz}(\omega) \l_{xx} (\omega)E_{z}^{\omega}E_{x}^{\omega}  \nonumber\\
 \mu _y^{2\omega }  &=& \l_{yy}(2\omega)\beta _{yyz} \l_{yy}(\omega)\l_{zz}(\omega)E_{y}^{\omega}  E_{z}^{\omega}    + \l_{yy}(2\omega)\beta _{yzy} \l_{zz}(\omega)\l_{yy}(\omega)E_{z}^{\omega}E_{y}^{\omega}  \nonumber\\
 \mu _z^{2\omega }  &=& \l_{zz}(2\omega)\beta _{zxx} \l_{xx} (\omega)\l_{xx}(\omega)E_{x}^{\omega}E_{x}^{\omega}    + \l_{zz}(2\omega)\beta _{zyy} \l_{yy}(\omega)\l_{yy}(\omega)E_{y}^{\omega} E_{y}^{\omega}  \nonumber\\
 && + \l_{zz}(2\omega)\beta _{zzz}
 \l_{zz}(\omega)\l_{zz}(\omega)E_{z}^{\omega}E_{z}^{\omega} \label{polarization-tensor}
\end{eqnarray}

Here the $E_{i}^{\omega}$ is from the Eq.\ref{MacroLocalfeil}, and
$\l_{ii}$ is the microscopic local field factors for the isotropic
monolayer using the discrete induced dipole model. $\l_{ii}$ is
explicitly derived and is defined as below.\cite{YePRB19834288}

\begin{eqnarray}
\l_{xx}(\omega_{i})&=& \bigg[1+\frac{\alpha _{xx}^{\omega
_{i}}}{2a^3}\xi_{0}+ \sum_{l=-\infty}^{\infty}{'\ \ }\frac{\alpha
_{xx}^{l,\omega
_{i}}}{2a_{l}^3}\xi^{l}_{0} \bigg]^{-1}\nonumber \\
\l_{yy}(\omega_{i})&=& \bigg[1+\frac{\alpha _{yy}^{\omega
_{i}}}{2a^3}\xi_{0}+ \sum_{l=-\infty}^{\infty}{'\ \ }\frac{\alpha
_{yy}^{l,\omega
_{i}}}{2a_{l}^3}\xi^{l}_{0} \bigg]^{-1}\nonumber \\
\l_{zz}(\omega_{i})&=& \bigg[1-\frac{\alpha
_{zz}^{\omega_{i}}}{a^3}\xi_{0} -\sum_{l=-\infty}^{\infty}{'\ \
}\frac{\alpha _{zz}^{l,\omega_{i}}}{a_{l}^3}\xi^{l}_{0}\bigg]^{-1}
\label{Localfieldfactor}
\end{eqnarray}

Here $\omega_{i}=\omega$ or $2\omega$, and $\alpha
_{ii}^{\omega_{i}}$ is the linear polarizability tensor which can be
calculated from the molecular polarizability values in the molecular
coordinates frame $\lambda'(a,b,c)$. The summation term in the
Eq.\ref{Localfieldfactor} is the dipole interaction from layers
other than the interface layer, which are generally negligible.
However, in the Section \ref{SectionIV}, the form shall be used in
the discussion of the segmentation of the chain molecules in the
molecular monolayer. The $\xi _0=-9.0336$ is for the square lattice
model as derived by Topping.\cite{ToppingJPRSLA1927} Since $\xi_{0}$
is negative, $l_{xx}=l_{yy}$ is generally larger than unity, while
$l_{zz}$ is smaller than unity. However, when $a$ is small, $|\alpha
_{xx}^{\omega _{i}}\xi_{0}/{2a^3}| > 1 $ can happen, this would
result in negative $l_{xx}=l_{yy}$ values even under the normal
dispersion condition for the bulk material. This is unique for the
two-dimensional cases.

The molecular polarizability at the experimental coordinates can be
expressed as from the coordinates transition relation: $ \alpha
_{ij}  = \sum\limits_{i' ,j'  = a,b,c} { {R_{ii' } R_{jj' } }
 \alpha _{i'j' } } $,  with $R_{\lambda,\lambda'}$ as the element of
the rotational transformation matrix from the molecular coordination
$\lambda'(a,b,c)$ to the laboratory coordination $\lambda(x,y,z)$.
\cite{HiroseApllSpec1051,WangHFIRPC2005} For the ensemble of
rotationally isotropic monolayer in the x-y plane, each individual
molecule possesses the same homogeneous distribution in the
azimuthal orientation and the twist orientation. Then, one has
\begin{eqnarray}
 \alpha _{xx}  &=& \frac{1}{4} \cdot \left( {1 + \cos ^2 \theta } \right) \cdot \alpha _{aa}  + \frac{1}{4} \cdot \left( {1 + \cos ^2 \theta } \right) \cdot \alpha _{bb}  + \frac{1}{2} \cdot \sin ^2 \theta  \cdot \alpha _{cc} \nonumber \\
 \alpha _{yy}  &=& \frac{1}{4} \cdot \left( {1 + \cos ^2 \theta } \right) \cdot \alpha _{aa}  + \frac{1}{4} \cdot \left( {1 + \cos ^2 \theta } \right) \cdot \alpha _{bb}  + \frac{1}{2} \cdot \sin ^2 \theta  \cdot \alpha _{cc} \nonumber \\
 \alpha _{zz}  &=& \frac{1}{2} \cdot \sin ^2 \theta  \cdot \alpha _{aa}  + \frac{1}{2} \cdot \sin ^2 \theta  \cdot \alpha _{bb}  + \cos ^2 \theta  \cdot \alpha
 _{cc}\label{alphaexpression}
\end{eqnarray}

Here we did not consider the case when the tilt angle $\theta$ is
with a distribution. In that case, the Eq.\ref{alphaexpression} need
to be put into the form with ensemble average over the distribution
of $\theta$. In doing calculations, the polarizability of the
molecule or molecular groups can be obtained from many compiled
sources,\cite{MillerJACS19797206,MillerJACS19908533,MillerJACS19908543}
or from direct quantum mechanics calculations.

The local field corrected linear and nonlinear polarizabilities in
the Eq.\ref{polarization-tensor} are the source of the linear or
nonlinear radiations from the monolayer at the interface. Their
detail expressions will not appear in the next section when we try
to calculate the total radiation from the monolayer. The detailed
expression in the Eq.\ref{polarization-tensor}, the
Eq.\ref{Localfieldfactor}, and the Eq.\ref{alphaexpression} can be
used for direct calculation of the local field factors and the local
field corrected polarizabilities in general.

The expression in the Eq.\ref{Localfieldfactor}, can be slightly
different if a non-square lattice model is assumed. For other
geometries, for example, the hexagonal or equi-triangular geometry,
the calculation of different geometries can be put forward according
to Topping's treatment.\cite{ToppingJPRSLA1927}

The Eq.\ref{Localfieldfactor} clearly indicates that when the
distance $a$ between the neighboring molecules in the monolayer
becomes large, i.e. the submonolayer case, the $\l_{ii}$ value is
approaching unity very quickly. This actually defines the meaning of
the word 'local', by considering how rapid the value $\xi_{0}$ is
converged when doing summation of the dipoles in the lattice plane
over $m$ and $n$. According to Topping and Philpott's calculations,
the convergence is generally reached in the 4th digit before $m,n <
10$.\cite{ToppingJPRSLA1927,PhilpottJCP1973595} For a lattice
constant of $a=5{\AA}$, this means the local field calculation
converges within 5nm. Since when $a$ become large, the local filed
factors in the Eq.\ref{Localfieldfactor} decays to unity rapidly,
therefore, the local field can be viewed as localized interactions
with few nanometers. The planewise sum rule also restricts the local
filed effect within only a few Angstroms from the dipole under
consideration. These facts provide definition of the actual range of
the local field effects.

By introducing the microscopic local field factors, the calculation
of the response from an ensemble of induced dipoles to an
electromagnetic field is reduced to a calculation of the response of
the isolated molecule interacting with the local field.

\section{Second Harmonic Radiation from the discrete interface induced dipoles in the far
field}\label{SectionIII}

Here we present the calculate of the linear and second harmonic
radiation in the far field according to the two-dimensional lattice
of the local field corrected induced dipoles at an interface between
the two isotropic bulk phases. The calculation using the square
lattice model can also be performed with hexagonal or other lattice
models. We shall see that since the radiation in the far field is
additive for the radiating dipoles, ensemble average treatment of
the radiating dipole can be straightforwardly implemented. The
following derivation uses the CGS convention, unless specified
otherwise.

In this problem, the radiation from the point induced dipole
$\vec{\mu}$ at the linear and second harmonic frequencies is in the
upper phase ($\varepsilon_{1}$) in
Fig.\ref{film-point-dipole-model1}. Here the point of origin
$O(0,0,0)$ is at the center of the dipole, different from the
definition in the previous section. This is for the convenience of
calculation in this section. Now, the general field radiated from an
electric dipole $\mu$ along the $z$ direction, i.e. $\mu \hat{z}$,
in the $\varepsilon_{1}$ phase is\cite{Bookmaxborn1997}

\begin{eqnarray}
 E_r  &=& \frac{{2\mu k^3_{1} }}{{\varepsilon _1 }}\left\{ {\frac{1}{{\left( {k_{1}r} \right)^3 }} - \frac{i}{{\left( {k_{1}r} \right)^2 }}} \right\}\cos \theta e^{i\left( {\vec{k_{1}}\cdot \vec{r} - \omega t} \right)} \nonumber \\
 E_\theta   &=& \frac{{\mu k^3_{1} }}{{\varepsilon _1 }}\left\{ {\frac{1}{{\left( {k_{1}r} \right)^3 }} - \frac{i}{{\left( {k_{1}r} \right)^2 }} - \frac{1}{{\left( {k_{1}r} \right)}}} \right\}\sin \theta e^{i\left( {\vec{k_{1}}\cdot \vec{r} - \omega t} \right)}\nonumber \\
 E_\varphi  &=& 0\label{radiationfield}
\end{eqnarray}

Here $\mu$ is the modulus of the induced-dipole, $k_{1}$ is the wave
vector in phase 1, $r$ is the distance from the point of the dipole
to the point of detection at $M(r,\theta,\varphi)$ with $\theta$ as
the tilt angle and $\varphi$ as the azimuthal angle in the polar
coordinates. If we only consider the far field, which is generally
in the experimental measurements except for the near field studies,
all the high order $1/r$ terms can be neglected. Therefore, only the
$1/r$ term of the $E_\theta$ in the Eq.\ref{radiationfield} need to
be considered.

In the following, the total macroscopic radiation field at a space
point $R(0,0, R_{0})$ in the reflection direction can be calculated
from the summation of the field of the radiation directly from the
individual induced dipole at the interface and the radiated field
reflected from the interface. In doing so, the amplitude and phase
of the radiated field at $R(0,0, R_{0})$ need to be expressed into
the function of the components of individual induced dipole at the
interface, before the summation can be carried out over them.

\begin{center}
\begin{figure}
\includegraphics[width=7cm]{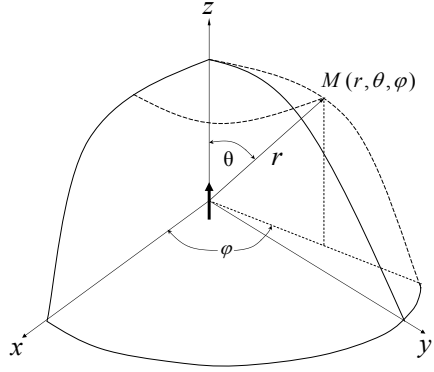} \caption{Polar coordinates with an electric dipole
with moment vector along the z-axis. $\mu$ is the modulus of the
induced-dipole, $r$ is the distance from the point of the dipole to
the point of detection, $\theta$ is the tilt angle in the polar
coordinates, $\varphi$ is the azimuthal angle.}\label{feild of
linear dipole}
\end{figure}
\end{center}

\begin{center}
\begin{figure}
\includegraphics[width=14cm,height=8cm]{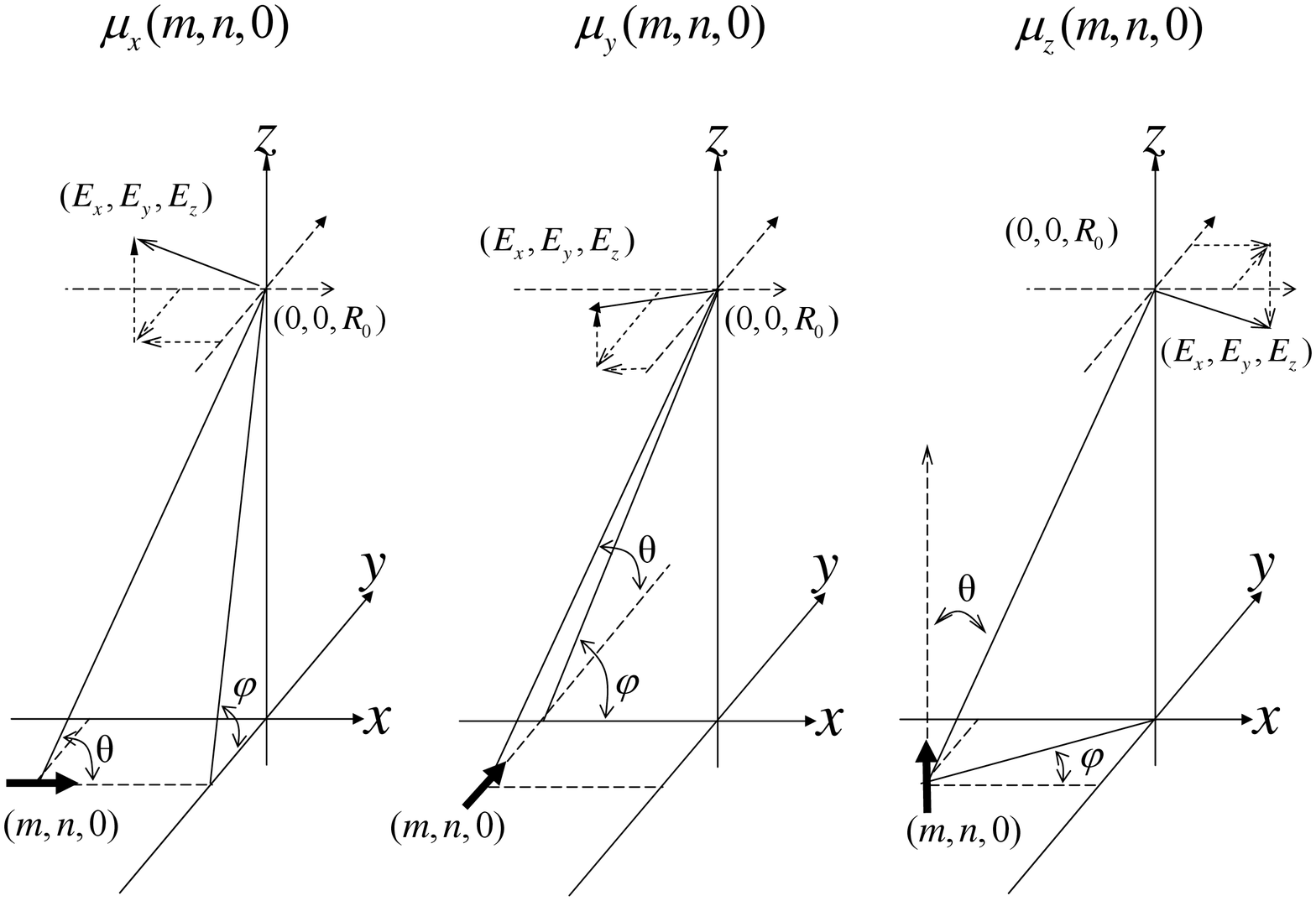} \caption{Calculation of the field vector ($E_{x}$,$E_{y}$,$E_{z}$),
which is created by $\mu_{x}(m,n,0)$, $\mu_{y}(m,n,0)$,
$\mu_{z}(m,n,0)$ respectively. $\theta$ being the tilt angle of the
dipole moment vector, $\varphi$ being the azimuthal angle of $r$
which is the vector from the point of (m,n,0) to the point
(0,0,$R_{0}$), as illustrated in Fig.\ref{feild of linear dipole}.
}\label{fieldofuxyz}
\end{figure}
\end{center}

The connection between the induced-dipole components $\mu_{x}$,
$\mu_{y}$ and $\mu_{z}$ and their radiation field components
$E_{x}$, $E_{y}$ and $E_{z}$ at $R(0,0,R_{0})$ is illustrated in
Figure \ref{fieldofuxyz}. The total phase at $R(0,0,R_{0})$ can be
calculated by choosing the point of origin $O(0,0,0)$ as the
reference point. Therefore the phase difference of the radiation
field at $R(0,0,R_{0})$ between the dipole at the point $r(ma,na,0)$
and the fixed point of origin $O(0,0,0)$ can be calculated from the
difference of the two distances $RO$ and $Rr$, i.e., $r(m,n)-R_{0}$,
with

\begin{equation}
r(m,n) = \sqrt {\left( {ma} \right)^2  + \left( {na} \right)^2  +
\left( {R_0 } \right)^2 } . \label{distance}
\end{equation}

Now, the phase difference of the incident field at $O$ and $r$ on
the dipole plane also need to be calculated. When a plane wave is
incident on the monolayer at the incident angle $\Omega_{i}$,
remembering that the incident plane is $xz$, and $m$ is along the
$x$ direction, this phase difference is $k^{in}_{1}(ma\sin
\Omega_{i} )$. For the case of second harmonic generation, this
phase difference involves two incoming waves, therefore the total
phase difference is $2k^{in}_{1}(ma\sin \Omega_{i})$.

Therefore, the total phase at the space point $R(0,0,R_{0})$ is the
sum of all these phase differences. If we define this total phase of
the radiation from the induced dipole at $r(ma,na,0)$ at
$R(0,0,R_{0})$, as $k_{1}f(m,n)$, one has

\begin{equation}
f(m,n) = \sqrt {\left( {ma} \right)^2  + \left( {na} \right)^2  +
\left( {R_0 } \right)^2 }  + \frac{\eta k^{in}_{1}}{k_{1}}(ma\sin
\Omega_{i} ). \label{phasedifference}
\end{equation}

Here when the radiation frequency is the same as the incoming
frequency $\omega$, then $\eta=1$, and $k_{1}^{in}=k_{1}$; while
when the radiation frequency is $2\omega$, then $\eta=2$, and
$k_{1}^{in}\neq k_{1}$. More generally, for the case of the sum
frequency generation ($\omega_{3}=\omega_{1} + \omega_{2}$), the
expression of the second term in $f(m,n)$ becomes

\begin{equation}
\frac{ma}{k_{1}(\omega_{3})}
[k_{1}^{in}(\omega_{1})sin\Omega_{i}^{\omega_{1}}+
k_{1}^{in}(\omega_{2})sin\Omega_{i}^{\omega_{2}}]
\label{phasedifferenceSFG}
\end{equation}

With the phase known, the radiation field at $R(0,0,R_{0})$
generated from the dipole $\vec{\mu}(\mu_{x},\mu_{y},\mu_{z})$ at
$r(ma,na,0)$ can be calculated separately for each $\mu_{i}$
according to the Eq.\ref{radiationfield} with proper projection.

Then the field components $E_{x}$, $E_{y}$,$E_{z}$ at $R(0,0,R_{0})$
generated by $\mu_{x}$ at $r(ma,na,0)$ are

\begin{eqnarray}
 E_{\mu _x ,x}  &=&  - E_{\mu _x ,\theta } \sin \theta  = \frac{{k^2_{1} \sin ^2 \theta }}{{\varepsilon_{1} r(m,n)}}\mu _{x}e^{i(k_{1}f(m,n) - \omega t)} \nonumber \\
 E_{\mu _x ,y}  &=& E_{\mu _x ,\theta } \cos \theta \cos \varphi  = -\frac{{k^2_{1} \sin \theta \cos \theta \cos \varphi }}{{\varepsilon_{1} r(m,n)}}\mu _{x}e^{i(k_{1}f(m,n) - \omega t)}  \nonumber\\
 E_{\mu _x ,z}  &=& E_{\mu _x ,\theta } \cos \theta \sin \varphi  = -\frac{{k^2_{1} \sin \theta \cos \theta \sin \varphi }}{{\varepsilon_{1} r(m,n)}}\mu _{x}e^{i(k_{1}f(m,n) - \omega t)} \label{fieldofux}
\end{eqnarray}
\noindent with
\begin{eqnarray}
 \sin \theta  &=& \frac{{\sqrt {(na)^2  + R_0 ^2 } }}{{\sqrt {(ma)^2  + (na)^2  + R_0 ^2 }
 }}~
 ,~
 \cos \theta  = \frac{{ - ma}}{{\sqrt {(ma)^2  + (na)^2  + R_0 ^2 } }} \nonumber\\
 \sin \varphi  &=& \frac{{R_0 }}{{\sqrt {(na)^2  + R_0 ^2 } }} ~,~
\cos \varphi  = \frac{{ - na}}{{\sqrt {(na)^2  + R_0 ^2 } }}
\end{eqnarray}

Similarly the field components $E_{x}$, $E_{y}$,$E_{z}$ at
$R(0,0,R_{0})$ generated by $\mu_{y}$ at $r(ma,na,0)$ are

\begin{eqnarray}
 E_{\mu _y ,x}  &=& E_{\mu _y ,\theta } \cos \theta \cos \varphi  =  - \frac{{k^2_{1} \sin \theta \cos \theta \cos \varphi }}{{\varepsilon_{1} r(m,n)}}\mu _{y} e^{i(k_{1}f(m,n) - \omega t)}  \nonumber\\
 E_{\mu _y ,y}  &=&  - E_{\mu _x ,\theta } \sin \theta  = \frac{{k^2_{1}\sin ^2 \theta }}{{\varepsilon_{1} r(m,n)}}\mu _{y} e^{i(k_{1}f(m,n) - \omega t)}  \nonumber\\
 E_{\mu _y ,z}  &=& E_{\mu _y ,\theta } \cos \theta \sin \varphi  =  - \frac{{k^2_{1} \sin \theta \cos \theta \sin \varphi }}{{\varepsilon_{1} r(m,n)}}\mu _{y} e^{i(k_{1}f(m,n) - \omega t)}\label{fieldofuy}
\end{eqnarray}
\noindent with
\begin{eqnarray}
 \sin \theta  &=& \frac{{\sqrt {(ma)^2  + R_0 ^2 } }}{{\sqrt {(ma)^2  + (na)^2  + R_0 ^2 } }}~,~\cos \theta  = \frac{{ - na}}{{\sqrt {(ma)^2  + (na)^2  + R_0 ^2 } }} \nonumber\\
 \sin \varphi &=& \frac{{R_0 }}{{\sqrt {(ma)^2  + R_0 ^2 } }}~,~\cos\varphi  = \frac{{ - ma}}{{\sqrt {(ma)^2  + R_0 ^2 }
 }}
\end{eqnarray}

Then the field components $E_{x}$, $E_{y}$,$E_{z}$ at $R(0,0,R_{0})$
generated by $\mu_{z}$ at $r(ma,na,0)$ are

\begin{eqnarray}
 E_{\mu _z ,x}  &=& E_{\mu _z ,\theta } \cos \theta \cos \varphi  =  - \frac{{k^2_{1} \sin \theta \cos \theta \cos \varphi }}{{\varepsilon_{1} r(m,n)}}\mu _{z} e^{i(k_{1}f(m,n) - \omega t)}  \nonumber\\
 E_{\mu _z ,y}  &=& E_{\mu _z ,\theta } \cos \theta \sin \varphi  =  - \frac{{k^2_{1} \sin \theta \cos \theta \sin \varphi }}{{\varepsilon_{1} r(m,n)}}\mu _{z} e^{i(k_{1}f(m,n) - \omega t)}  \nonumber\\
 E_{\mu _z ,z}  &=&  - E_{\mu _z ,\theta } \sin \theta  = \frac{{k^2_{1} \sin ^2 \theta }}{{\varepsilon_{1} r(m,n)}}\mu _{z} e^{i(k_{1}f(m,n) - \omega t)}\label{fieldofuz}
\end{eqnarray}
\noindent with
\begin{eqnarray}
 \sin \theta  &=& \frac{{\sqrt {(ma)^2  + (na)^2 } }}{{\sqrt {(ma)^2  + (na)^2  + R_0 ^2 } }}~,~\cos \theta  = \frac{{R_0 }}{{\sqrt {(ma)^2  + (na)^2  + R_0 ^2 } }} \nonumber\\
 \sin \varphi  &=& \frac{{ - na}}{{\sqrt {(ma)^2  + (na)^2 } }}~,~\cos \varphi  = \frac{{ - ma}}{{\sqrt {(ma)^2  + (na)^2 }
 }}
\end{eqnarray}

Now the total radiation field at $R(0,0,R_{0})$ generated by the
whole monolayer at the interface is the summation over the whole
lattice plane $m$(-$\infty$,+$\infty$)~and $n$~$(-\infty,+\infty)$.

\begin{eqnarray}
 E_x  = \sum\limits_{m =  - \infty }^\infty  {\sum\limits_{n =  - \infty }^\infty  {\left( {E_{\mu _x ,x}  + E_{\mu _y ,x}  + E_{\mu _z ,x} } \right)} }  \nonumber\\
 E_y  = \sum\limits_{m =  - \infty }^\infty  {\sum\limits_{n =  - \infty }^\infty  {\left( {E_{\mu _x ,y}  + E_{\mu _y ,y}  + E_{\mu _z ,y} } \right)} }  \nonumber\\
 E_z  = \sum\limits_{m =  - \infty }^\infty  {\sum\limits_{n =  - \infty }^\infty  {\left( {E_{\mu _x ,z}  + E_{\mu _y ,z}  + E_{\mu _z ,z} } \right)} }
 \label{totallradiation}
\end{eqnarray}

The summation in the Eq.\ref{totallradiation} for the total
radiation field is in principle the discrete form of the
integro-differential equation in the molecular
optics.\cite{Bookmaxborn1997,WolfJOSA1972} The asymptotic result of
this summation can be evaluated by employing the condition of the
far field. Because $R_{0}$ is very large as compared with the
lattice constant $a$ and the wavelength $\lambda$, the discrete
summations in the Eq.\ref{totallradiation} is asymptotic to a
continuous integration over the variable set $(x, y)=(ma, na)$. In
addition, because $R_{0}$ is large and so the phase factor
$k_{1}f(m,n)$ is large, the exponential factors in the
Eq.\ref{fieldofux}, \ref{fieldofuy} and \ref{fieldofuz} oscillate
very rapidly and change their signs many times as the point $r(ma,
na, 0)$ explores the domain of integration.\cite{Bookmaxborn1997}
Under these conditions, an asymptotic value of the total electric
field ($E_x,E_y,E_z$) at point $R(0,0,R_{0})$ can be obtained with
the application of the following formula, which are derived from the
principle of stationary phase. The detail of the stationary phase
method can be found in classic textbook by Born and
Wolf.\cite{Bookmaxborn1997}

Here by defining the variables $x=ma$ and $y=na$ with the lattice
constant $a$ as a very small quantity compared to the infinity size
of the lattice plane, each term in the summation as shown in the
Eq.\ref{totallradiation} becomes the following integral and this
integral is then asymptotic to the summation of the stationary phase
terms as below.

\begin{eqnarray}
\frac{1}{a^{2}}\int^{\infty}_{-\infty} {\int^{\infty}_{-\infty}
{g\left( {x,y} \right)} } e^{ik_{1}f\left( {x,y} \right)} dxdy =
\frac{{2\pi i}}{k_{1}a^{2}}\sum\limits_j {\frac{{\sigma _j }}{{\sqrt
{\left| {\alpha _j \beta _j  - \gamma _j ^2 } \right|} }}} g\left(
{x_j ,y_j } \right)e^{ik_{1}f\left( {x_j ,y_j } \right)}
\label{approximation}
\end{eqnarray}

Here $g\left({x,y} \right)$ is the pre-exponential factors in the
Eq.\ref{fieldofux}, \ref{fieldofuy} and \ref{fieldofuz}. $(x_j
,y_j)$ is the points in the whole integration domain at which $f$ is
stationary, i.e.

\begin{eqnarray}
\frac{{\partial f}}{{\partial x}} = \frac{{\partial f}}{{\partial
y}} = 0 \label{StationaryCondition}
\end{eqnarray}

The definition of the other terms are the followings.

\begin{eqnarray}
\alpha _j  = \left( {\frac{{\partial ^2 f}}{{\partial x^2 }}}
\right)_{x_j ,y_j } ,\beta _j  = \left( {\frac{{\partial ^2
f}}{{\partial y^2 }}} \right)_{x_j ,y_j } ,\gamma _j  = \left(
{\frac{{\partial ^2 f}}{{\partial x\partial y}}} \right)_{x_j ,y_j }
\end{eqnarray}
and
\begin{eqnarray}
\sigma _j  = \begin{array}{*{20}c}
   { + 1~~~~~~if~~\alpha _j \beta _j  > \gamma _j ^2 ~,~~\alpha _j  > 0~,}  \\
   { - 1~~~~~~if~~\alpha _j \beta _j  > \gamma _j ^2 ~,~~\alpha _j  < 0~,}  \\
   { - i~~~~~~if~~\alpha _j \beta _j  < \gamma _j ^2 ~.~~~~\begin{array}{*{20}c}
   {} & {} & {} & {} & {}\\
\end{array}}
\end{array}
\end{eqnarray}

Using the stationary phase condition in the
Eq.\ref{StationaryCondition} and the expression of $f$ in the
Eq.\ref{phasedifference}, a single stationary point $(x_{1}, y_{1})$
is obtained, and

\begin{eqnarray}
x_1  = -R_{0} \tan \Omega ~~,~~y_1  = 0~~.
\end{eqnarray}

Here $\Omega=\Omega_{i}$ is the direct results of the stationary
phase condition for the linear radiation, and for the second
harmonic radiation $\Omega$ has to satisfy the relationship
$k_{1}(2\omega)\sin\Omega=2k_{1}^{in}(\omega)\sin\Omega_{i}$.
Similarly, one can also show explicitly that for the sum frequency
radiation, $\Omega$ has to satisfy the relationship
$k_{1}(\omega_{3})\sin\Omega=k_{1}^{in}(\omega_{1})\sin\Omega_{i}^{\omega_{1}}+
k_{1}^{in}(\omega_{2})\sin\Omega_{i}^{\omega_{2}}$. These are just
the condition for the general law of reflection in linear and
nonlinear optics, as given by Bloembergen and Pershan, as derived
from the Maxwell boundary conditions.\cite{BloembergenPR1962606}
From these results we shall show later that at the far field space
point $R(0,0,R_{0})$, the radiation field is in the direction of
$\Omega$, which satisfies the general law of reflection in linear
and nonlinear optics. It is interesting that these conditions came
incidentally as a solution from the molecular optics treatment. This
fact further illustrates the general equivalency of the macroscopic
Maxwell equations with boundary conditions and the molecular optics
treatment.\cite{WolfJOSA1972,Bookmaxborn1997}

Now with the stationary phase point value, we have
\begin{eqnarray}
 f\left( {x_1 ,y_1 } \right) = R_0 \cos \Omega  \nonumber\\
 \frac{{\sigma _1 }}{{\sqrt {\left| {\alpha _1 \beta _1  - \gamma _1 ^2 } \right|} }} = \frac{{R_0 }}{{\cos ^2 \Omega}}
 ~~.
\end{eqnarray}

Here $\sigma_{1}=1$ because $\alpha_{1}>0$ and
$\alpha_{1}\beta_{1}-\gamma_{1}^{2}> 0$. Now put all these values
into the expressions in the Eq.\ref{totallradiation}, we have the
radiation field components at the space point $R(0,0,R_{0})$ that
are directly radiated from the whole induced dipole lattice as the
followings.

\begin{eqnarray}
 E^{R_{1}}_x  &=& \frac{2\pi}{{a^2 \varepsilon_{1}}}ik_{1}\left( {\mu _x \cos \Omega  - \mu _z \sin \Omega } \right)e^{i(k_{1}R_0 \cos \Omega -\omega t)}\nonumber\\
 E^{R_{1}}_y  &=& \frac{2\pi}{{a^2 \varepsilon_{1}}}ik_{1}\frac{{\mu _y }}{{\cos \Omega }} e^{i(k_{1}R_0 \cos \Omega -\omega t)}\nonumber\\
 E^{R_{1}}_z  &=& \frac{2\pi}{{a^2 \varepsilon_{1}}}ik_{1}\left( {-\mu _x \sin \Omega  + \mu _z \frac{{\sin ^2 \Omega }}{{\cos \Omega}}}\right)e^{i(k_{1}R_0 \cos \Omega -\omega t)}
 \label{reflectoftotallradiation}
\end{eqnarray}

In order to calculate the total radiated field at $R(0,0,R_{0})$,
besides the direct radiation from the induced dipoles at the
interface, the radiation in the forward direction which is reflected
from the interface also need to be evaluated. The above procedures
can be repeated to calculate the radiated field at the image point
$R_{i}(0,0,-R_{0})$ in the isotropic phase $\varepsilon_{1}$ from
the whole induced dipole lattice. We have

\begin{eqnarray}
 E^{T_{1}}_x  &=& \frac{2\pi}{{a^2 \varepsilon_{1}}}ik_{1}\left( {\mu _x \cos \Omega  + \mu _z \sin \Omega} \right)e^{i(k_{1}R_0 \cos \Omega -\omega t)}\nonumber\\
 E^{T_{1}}_y  &=& \frac{2\pi}{{a^2 \varepsilon_{1}}}ik_{1}\frac{{\mu _y }}{{\cos \Omega }}e^{i(k_{1}R_0 \cos \Omega -\omega t)}\nonumber\\
 E^{T_{1}}_z  &=& \frac{2\pi}{{a^2 \varepsilon_{1}}}ik_{1}\left( {\mu _x \sin \Omega  + \mu _z \frac{{\sin ^2 \Omega }}{{\cos \Omega }}}\right)e^{i(k_{1}R_0 \cos \Omega -\omega t)}
 \label{refractoftotallradiation}
\end{eqnarray}

Because the choice of the space point $R(0,0,R_{0})$ and
$R_{i}(0,0,-R_{0})$ is rather arbitrary, and the radiation field
components in the Eq.\ref{reflectoftotallradiation} and the
Eq.\ref{refractoftotallradiation} are both directional and the
amplitudes of both plane waves are independent from the value of
$R_{0}$, the coherence of the radiation from the induced dipole
lattice is conserved at the far field.

According to the Fresnel formulae,\cite{Bookmaxborn1997} the
amplitudes of the reflected electric field components are
\begin{eqnarray}
 E_x^{R_2 }  &=& -r_{p}E^{T_{1}}_x=\frac{{n_1 \cos \Omega_t  - n_2 \cos \Omega_i }}{{n_2 \cos \Omega_i  + n_1 \cos \Omega_t }}E^{T_{1}}_x  \nonumber\\
 E_y^{R_2 }  &=& r_{s}E^{T_{1}}_y=\frac{{n_1 \cos \Omega_i  - n_2 \cos \Omega_t }}{{n_1 \cos \Omega_i  + n_2 \cos \Omega_t }}E^{T_{1}}_y  \nonumber\\
 E_z^{R_2 }  &=& r_{p}E^{T_{1}}_z=\frac{{n_2 \cos \Omega_i  - n_1 \cos \Omega_t }}{{n_2 \cos \Omega_i  + n_1 \cos \Omega_t
 }}E^{T_{1}}_z \label{relfection2}
\end{eqnarray}

The phase difference between the direct radiation field and the
reflected field is determined by the distance between the dipole
layer and the interface $d$ and the radiation angle $\Omega$ as
$2k_{1}d\cos \Omega$. Therefore the total radiation field at
$R(0,0,R_{0})$ is the phase shifted sum of the $E^{R_{1}}$ and
$E^{R_{2}}$ fields in the Eq.\ref{reflectoftotallradiation} and the
Eq.\ref{relfection2}. We have

\begin{eqnarray}
 E_x,total   &=& \frac{2\pi}{{a^2 \varepsilon_{1}}}ik_{1}\left( {\mu _{x} \cos \Omega  - \mu _{z} \sin \Omega } -r_{2p}({\mu _{x} \cos \Omega  + \mu _{z} \sin \Omega })e^{i2k_{1}d\cos \Omega } \right )e^{i(k_{1}R_{0}\cos \Omega -\omega t)}\nonumber\\
              &=& \frac{2\pi}{{a^2 \varepsilon_{1}}}ik_{1}\left( \cos \Omega L'_{xx}{\mu_{x}  - \sin \Omega L'_{zz}\mu_{z}  } \right )e^{i(k_{1}R_{0}\cos \Omega -\omega t)}\nonumber\\
 E_y,total   &=& \frac{2\pi}{{a^2 \varepsilon_{1}}}ik_{1}\left( \frac{{\mu _{y} }}{{\cos \Omega }} + r_{2s}\frac{{\mu _{y} }}{{\cos \Omega }}e^{i2k_{1}d\cos \Omega }\right )e^{i(k_{1}R_{0}\cos \Omega -\omega t)}\nonumber\\
              &=& \frac{2\pi}{{a^2 \varepsilon_{1}}}ik_{1}\left( \frac{{1}}{{\cos \Omega }}L'_{zz}{\mu_{y} }\right )e^{i(k_{1}R_{0}\cos \Omega - \omega t)}\nonumber\\
 E_z,total   &=& \frac{2\pi}{{a^2 \varepsilon_{1}}}ik_{1}\left( {-\mu _{x} \sin \Omega  + \mu _{z} \frac{{\sin ^2 \Omega }}{{\cos \Omega }}} +r_{2s}({\mu _{x} \sin \Omega  + \mu _{z} \frac{{\sin ^2 \Omega }}{{\cos \Omega}}} )e^{i2k_{1}d\cos \Omega }\right)e^{i(k_{1}R_{0}\cos \Omega -\omega t)}\nonumber \\
              &=& \frac{2\pi}{{a^2 \varepsilon_{1}}}ik_{1}\left( -\sin \Omega L'_{xx} \mu _{x} + \frac{{\sin ^2 \Omega}}{{\cos \Omega }}L'_{zz}\mu _{z} \right )e^{i(k_{1}R_{0}\cos \Omega -\omega t)}
 \label{Totalrefractoftotallradiation}
\end{eqnarray}

Here the induced dipole components $\mu_{x}, \mu_{y}, and \mu_{z}$
are as defined in the Eq.\ref{polarization-tensor}, and the
$L'_{ii}$ are as defined in the Eq.\ref{MacroLocalfeil}. Considering
the fact that the thickness of the monolayer is generally much
smaller than the magnitude of the optical wavelength, i.e. $d\ll
\lambda$, the phase factor $e^{i2k_{1}d\cos \Omega }\rightarrow 1$.
Therefore, $L'_{ii}$ becomes

\begin{eqnarray}
 L_{xx}  &=& \frac{{2n_1 \cos \Omega_t }}{{n_2 \cos \Omega_{i}  + n_1 \cos \Omega_t }}=\frac{{2\varepsilon_1 k_{2z}}}{{\varepsilon_2 k_{1z}  + \varepsilon_1 k_{2z} }}\nonumber \\
 L_{yy}  &=& \frac{{2n_1 \cos \Omega_{i} }}{{n_1 \cos \Omega_{i}  + n_2 \cos \Omega_t }}=\frac{{2k_{1z}}}{{k_{1z}  + k_{2z}}}\nonumber \\
 L_{zz}  &=& \frac{{2n_2 \cos \Omega_{i} }}{{n_2 \cos \Omega_{i}  + n_1 \cos \Omega_t }}=\frac{{2\varepsilon_2 k_{1z}}}{{\varepsilon_2 k_{1z}  + \varepsilon_1 k_{2z} }}\label{MacroLocalfeilapproximation}
\end{eqnarray}

These $L_{ii}$ values are clearly the results for the two phase
model. This indicates explicitly that when dealing with the
macroscopic electric field at the interface, the existence of the
interfacial molecular monolayer has minimum influence on the
macroscopic field. Therefore, the problem of the linear macroscopic
dielectric constant for the interface layer is not a issue in this
treatment. This shall be discussed in detail in the Section
\ref{SectionIV}.A.

Now the radiation field components in the second harmonic frequency
are

\begin{eqnarray}
 E_x^{2\omega}   &=& \frac{2\pi}{{a^2 \varepsilon_{1}(2\omega)}}ik^{2\omega}_{1}\left( \cos \Omega_{2\omega} L^{2\omega}_{xx}{\mu^{2\omega} _{x}   - \sin \Omega_{2\omega} L^{2\omega}_{zz}\mu^{2\omega} _{z}  } \right )e^{i(k^{2\omega}_{1}R_{0}\cos \Omega_{2\omega} -2\omega t)}\nonumber\\
 E_y^{2\omega}   &=& \frac{2\pi}{{a^2 \varepsilon_{1}(2\omega)}}ik^{2\omega}_{1}\left( \frac{{1}}{{\cos \Omega_{2\omega} }}L^{2\omega}_{zz}{\mu^{2\omega} _{y} }\right )e^{i(k^{2\omega}_{1}R_{0}\cos \Omega_{2\omega} -2\omega t)}\nonumber\\
 E_z^{2\omega}   &=& \frac{2\pi}{{a^2 \varepsilon_{1}(2\omega)}}ik^{2\omega}_{1}\left( -\sin \Omega_{2\omega} L^{2\omega}_{xx} \mu^{2\omega} _{x} + \frac{{\sin ^2 \Omega_{2\omega}}}{{\cos \Omega_{2\omega} }}L^{2\omega}_{zz}\mu^{2\omega} _{z} \right )e^{i(k^{2\omega}_{1}R_{0}\cos \Omega_{2\omega} -2\omega t)}\label{simplySHG}
\end{eqnarray}

Here $k^{2\omega}_{1}=2\omega \sqrt{\varepsilon_{1}(2\omega)}/c$,
where $c$ is the velocity of light in vacuum. Now with the second
harmonic field components determined, the intensity of the SH field
can be directly calculated by the magnitude of the time averaged
Poynting vector as defined below.\cite{BoydBook}

\begin{eqnarray}
 I(\omega)&=&\frac{c}{2\pi}\sqrt{\varepsilon_{1}(\omega)}\big|E_{i}(\omega)\big|^{2}\label{Poynting}
 \end{eqnarray}

Then put the expressions in the Eq.\ref{simplySHG} and the
Eq.\ref{polarization-tensor} into the Eq.\ref{Poynting}, and after
taking care of the proper projection coefficients of the fundamental
and second harmonic electric fields, we have

\begin{eqnarray}
 I(2\omega )&=&\frac{{32\pi ^3 \omega ^2 \sec ^2 \Omega_{2\omega} }}{{\textit{c} ^3[\epsilon_{1}(2\omega)]^{1/2}\epsilon_{1}(\omega)}}|\chi _{eff}(2\omega) |^2 I^{2}(\omega)\label{generaleffectivesusceptibility}\\
\chi_{eff}&=&[\mathbb{L}(2\omega)\cdot\hat{e}(2\omega)]\cdot\chi(2\omega):[\mathbb{L}(\omega)\cdot
\hat{e}(\omega)][\mathbb{L}(\omega)\cdot
\hat{e}(\omega)]\label{Chi1}
 \end{eqnarray}

\noindent Here $I(\omega)$ is the intensity of the incident laser
beam as defined in the Eq.\ref{Poynting}, and $\Omega_{2\omega}$ is
outgoing angle of the second harmonic radiation from the surface
normal. $\epsilon_{1}(\omega)$ and $\epsilon_{1}(2\omega)$ are the
bulk dielectric constants of the upper bulk phase at the frequency
$\omega$ and $2\omega$, respectively. $\hat{e}(2\omega )$ and
$\hat{e}(\omega )$ are the unit vectors of the electric field at
$2\omega$ and $\omega$, respectively. $\mathbb{L}(2\omega)$ and
$\mathbb{L}(\omega)$ are the tensorial Fresnel factors for $2\omega$
and $\omega$, respectively, as defined as in the
Eq.\ref{MacroLocalfeilapproximation}. The scaler property
$\chi_{eff}(2\omega)$ is the effective macroscopic second-order
susceptibility of the interface. The $\chi(2\omega)$ represents the
area-averaged macroscopic second harmonic susceptibility tensors
defined as
$\chi_{ijk}(2\omega)=l_{ii}(2\omega)\l_{jj}(\omega)\l_{kk}(\omega)\beta_{ijk}/a^{2}=N_{s}l_{ii}(2\omega)\l_{jj}(\omega)\l_{kk}(\omega)\langle\beta_{i'j'k'}\rangle$,
with $a$ as the lattice constant, $N_{s}$ as the surface number
density, and $\langle \  \rangle$ represents ensemble average over
different orientational distributions.

The expressions in the Eq.\ref{simplySHG} and the
Eq.\ref{generaleffectivesusceptibility} are exactly the same as
those in the previous treatment by Heinz and
Shen.\cite{ShenARPC1989,Heinzthesis1982,HeinzBook,ZhuangPRB199912632,ShenFellerPRA1991}
This indicates that the treatment with the square point-dipole
lattice model together with a molecular optics approach is at least
phenomenologically correct. It also further validates the treatment
in the original literatures.

It can be shown accordingly that for the SFG from the interface, one
has

\begin{eqnarray}
 I(\omega_{SF})&=&\frac{{8\pi ^3 \omega ^2 \sec ^2 \Omega_{SF} }}{{\textit{c} ^3[\epsilon_{1}(\omega_{SF})\epsilon_{1}(\omega_{1})\epsilon_{1}(\omega_{2})]^{1/2}}}\big|\chi _{eff}(\omega_{SF}) \big|^2 I(\omega_{1})I(\omega_{2})\label{SFGIntensity}\\
\chi_{eff}&=&[\mathbb{L}(\omega_{SF})\cdot\hat{e}(\omega_{SF})]\cdot\chi(\omega_{SF}):[\mathbb{L}(\omega_{1})\cdot
\hat{e}(\omega_{1})][\mathbb{L}(\omega_{2})\cdot
\hat{e}(\omega_{2})]\label{ChiSFG}
 \end{eqnarray}

Now, the following general issues need to be discussed.

a. Even though the expressions in the Eq.\ref{simplySHG} and the
Eq.\ref{generaleffectivesusceptibility} were obtained through the
square induced dipole lattice model, using other type of plane
lattice model gives the same asymptomatic expressions at the far
field, except that the area-averaged macroscopic second harmonic
susceptibility tensors take slightly different expressions.
Certainly, the near field results have to be different for different
lattice models, because generally no analytical asymptomatic results
can be reached for the near field case. In other words, the
molecular optics approach described here can be used to treat the
near field cases, this is an clear advantage of the molecular optics
approach.

b. Here the area-averaged macroscopic second harmonic susceptibility
tensors $\chi_{ijk}(2\omega)$ and the factor $1/a^{2}$ in the
Eq.\ref{simplySHG} needs to be discussed. $a^{2}$ is the average
area per induced dipole in the lattice plane under the square
induced-dipole lattice model. It comes into the expression naturally
because of the summation, or integration, over the whole lattice
plane. Therefore, $\mu_{i}/a^{2}$ is the area averaged induced
dipole, and $\chi_{ijk}(2\omega)$ is the area-averaged macroscopic
second harmonic susceptibility tensors. In the previous treatments
on second harmonic generation,\cite{Heinzthesis1982,ShenARPC1989} an
infinitesimally thin polarization sheet layer was used, and the
polarization for this interface layer was defined in the form of
$P(2\omega)=\chi(2\omega)\delta(z):E(\omega)E(\omega)$, or sometimes
$P(2\omega)=\chi(2\omega):E(\omega)E(\omega)$.\cite{Heinzthesis1982,ShenARPC1989,Brevetbook,ShenNonlinearOpticsBook}
In this way, it is a little confusing when one try to make the
dimension analysis of the formulae when trying to apply the
interface polarization term into the Maxwell equations. However, in
this molecular optics treatment, there is no ambiguity involved, and
the $\chi(2\omega)$ tensor has to have the dimension of an area
-averaged second order nonlinear polarizability.

c. The ensemble average of the $\chi_{ijk}(2\omega)$ tensors. In
calculation of the summation and its asymptomatic results in the far
field, the actual size of the lattice constant $a$ can be treated
arbitrarily as long as the lattice constant $a$ is microscopic.
Because the additive nature of the summation in the
Eq.\ref{totallradiation}, it is easy to show that even though the
surface is not a square lattice for each individual dipole, one can
always divide the surface layer into the square lattice with
microscopic unit cells each containing a group of individual
neighboring induced dipoles. Therefore, this approach actually
allows making ensemble average with each unit cell. It even allow
the unit cell to contain several layers of molecules close to the
interface region. This further allows the inclusion of the
contributions of the quadrupolar and the interfacial discontinuity
terms into the effective unit induced dipole cells at the interface
region.\cite{BloembergenPR1962606,
GuyotsionnestPRB19868254,ShenARPC1989,Brevetbook,ShenNonlinearOpticsBook,ZhangWK2005JCP123p224713}
Of course, each of these different treatments has to be involved
with different microscopic description and symmetry analysis of the
area-averaged effective microscopic polarizability tensor terms
$\chi_{ijk}$ of the unit cell as defined.

With above issues cleared, now we may say that the
Eq.\ref{simplySHG} and the Eq.\ref{generaleffectivesusceptibility}
are general for the second harmonic generation from the whole
interface region, with the possibility to make microscopic treatment
or calculation with different microscopic models of the interface
region. Similar treatment can be performed for the sum frequency
generation process from the interface. Its detail is not discussed
here.

In the SHG, most of the time a rotationally isotropic molecular
interface is studied. If we let $\alpha_{in}$ be the polarization
angle of the incident electric field, let $\gamma_{out}$ be the
polarization angle of the second harmonic electric field,
$\chi_{eff,\alpha_{in}-\gamma_{out}}$ be the corresponding effective
macroscopic nonlinear susceptibility. Then the
$\chi_{eff,\alpha_{in}-\gamma_{out}}$ can be expressed into the
linear combinations of the three independently measurable terms with
the following polarization combinations: $s_{in}-p_{out}$,
$45^{\circ}_{in}-s_{out}$, $p_{in}-p_{out}$, with $s$ as the
polarization perpendicular to the incident plane, and $p$ as the
polarization in the incident plane. Let $\Omega_{\omega}$ and
$\Omega_{2\omega}$ be the incoming and outgoing angles,
respectively, one has\cite{WangHFPCCP2006}

\begin{eqnarray}
\chi_{eff,\alpha_{in}-\gamma_{out}}&=&\chi
_{eff,45^{\circ}-s}\sin\gamma\sin2\alpha+\chi
_{eff,s-p}\cos\gamma\sin^{2}\alpha+\chi _{eff,p-p}\cos\gamma
\cos^{2}\alpha \label{SHGExperiment}
 \end{eqnarray}

 \begin{eqnarray}
\chi_{eff,s-p}  &=& L_{zz} (2\omega )L_{yy}
(\omega )L_{yy} (\omega )\sin \Omega_{2\omega} \chi _{zyy} \nonumber\\
\chi _{eff,45^{\circ} -s}  &=& L_{yy} (2\omega )L_{zz} (\omega
)L_{yy} (\omega )\sin \Omega_{\omega} \chi _{yzy} \nonumber\\
\chi _{eff,p-p}  &=& - 2L_{xx} (2\omega )L_{zz}
 (\omega )L_{xx} (\omega )\cos \Omega_{2\omega} \sin \Omega_{\omega} \cos \Omega_{\omega}  \chi _{xzx} \nonumber \\
  &&+ L_{zz} (2\omega )L_{xx} (\omega )L_{xx} (\omega )\sin \Omega_{2\omega} \cos ^2 \Omega_{\omega} \chi _{zxx} \nonumber \\
  &&+ L_{zz} (2\omega )L_{zz} (\omega )L_{zz} (\omega )\sin \Omega_{2\omega}\sin ^2 \Omega_{\omega}\chi _{zzz}
 \label{effectivesusceptibility}
\end{eqnarray}

Details of the quantitative experimental measurement and
interpretation of the interface SHG, as well as the SFG-vibrational
spectroscopy, of molecular interfaces with these formulations can be
found in the recent literatures.\cite{WangHFPCCP2006,WangHFIRPC2005}

\section{Discussion}\label{SectionIV}

Two of the central issues with general interests in the quantitative
interpretation of the SHG and SFG-VS data are whether the molecular
monolayer has a linear macroscopic dielectric constant and how the
microscopic local field effects is evaluated. Actually previous
researchers already provided most of the answers as reviewed in the
introduction section above. Because the SHG and the SFG-VS have been
proven as the sensitive probes for interfaces with the submonolayer
coverage,\cite{ShenNature1989,ShenARPC1989,Corn1994ChemRev,Eisenthal1996ChemRev}
the treatment based on the more realistic discrete induced dipole
lattice model is developed in this work. Based on the microscopic
treatment of the SHG and SFG-VS from the molecular interface, these
issues are to be discussed.

\subsection{The linear macroscopic dielectric constant of the molecular monolayer or submonolayer}

The concern of the linear macroscopic dielectric constant of the
interface is from the beginning of the development of SHG and SFG-VS
as the quantitative surface spectroscopic
probes.\cite{ShenNonlinearOpticsBook} Here we quote the following
sentences from the Chapter `Surface Nonlinear Optics' in Shen's
classical textbook:\cite{ShenNonlinearOpticsBook}

`In the usual simplified approach, the surface microscopic layer is
assumed to have a characterized thickness with optical constants
different from the bulk. Because of the linear transmission and
reflection of the light at the boundary surface usually are
dominated by the bulk properties, the surface layers have little
effect on the linear wave propagation. As far as nonlinear optical
effects are concerned, we can assume that the surface layer have the
same linear refractive indices as the adjoining bulk media, but
their nonlinear optical susceptibilities are different from the
bulk. Unlike the linear case, the surface layer can strongly affect
the nonlinear optical output in some cases.'

The insight in these words is the basis for the later treatment by
Shen to substitute the macroscopic dielectric constant of the
interface layer in the early
treatment\cite{Heinzthesis1982,ShenFellerPRA1991} with the
microscopic local field
factors.\cite{YePRB19834288,ZhuangPRB199912632,WeiXing2000PRE62p5160}
This implies that the linear macroscopic dielectric constant of the
interface layer does not need to be considered. Here the rigorous
molecular optics treatment in the Section \ref{sectionII} and
\ref{SectionIII} with the discrete induced dipole lattice model is
providing a further support for this approach.

Based on the following reasonings, the linear macroscopic dielectric
constant of the monolayer or submonolayer is not an issue in the
molecular optics approach described in the Section \ref{sectionII}
and \ref{SectionIII} explicitly.

a. Firstly, we consider a few nonlinear induced dipoles situated at
the interface between the two isotropic bulk phases, for example,
the vacuum/metal or other fluid/solid interfaces. These interface
induced dipoles are situated in the vacuum phase and they does not
even form an interface layer. Therefore, under the submonolayer
condition up to a full monolayer condition, the treatment as
presented in the Section \ref{sectionII} and \ref{SectionIII} with
the discrete induced dipole lattice model is rigorous, and no linear
macroscopic dielectric constant can be invoked for the molecular
layer. When this is validated, then consider the case that there is
a submonolayer up to a full monolayer on top of the first full
monolayer at the air/metal interface. The treatment of this top
layer must be the same, i.e., no need to invoke a linear macroscopic
dielectric constant for this top layer, and only the microscopic
local field effect, i.e. the dipole screening effect, need to be
considered.\cite{BagchiPRL19801475,YePRB19834288} Since the
interface layer in between any two isotropic bulk phases is usually
one or few molecular monolayer thick, the discrete induced dipole
lattice model must be a generally realistic treatment of the
interface problem. In simple words, this implies that there is no
such thing can be defined as the linear macroscopic dielectric
constant for the interface layer.

b. Secondly, let us consider the case for the submonolayer
adsorption of organic molecules at the air/liquid, e.g. submonolayer
of surfactant molecules at the air/water interface. Since the
air/water interface is known as sharp as less than
$4{\AA}$,\cite{RiceJCP1991,PershanPRL1985,PershanPRA1988} one can
surmise that the hydrophobic part is in the air phase and the
hydrophilic part is in the water phase. In this case, the treatment
in the Section \ref{sectionII} and \ref{SectionIII} with discrete
induced dipole lattice model is generally a realistic treatment of
the interface problem, especially for the SFG-VS when the tail and
the head group of the interfacial surfactant molecule are probed
with different vibrational frequencies. The radiation from the
molecular induced dipole in the phase 2 ($\varepsilon_{2}$) can also
be treated similarly and readily as in the Section \ref{sectionII}
and \ref{SectionIII}. Nevertheless, in these cases, the continuous
surface sheet layer model does not provide a realistic
representation of the microscopic molecular picture.

In the above cases, the linear macroscopic dielectric constant of
the interface layer should not be a concern for the surface
nonlinear optics whenever the surface contribution is dominant.
Therefore, the macroscopic optical field at the interface can be
well described with an explicit two phase model, as in the
Eq.\ref{MacroLocalfeilapproximation}. The major concern by
Roy\cite{RoyPRB200013283} on the macroscopic linear anisotropy of
the interface monolayer in the SHG treatment is not a real issue at
least for the submonolayer up to the monolayer
regime.\cite{ShenNonlinearOpticsBook} Nevertheless, according to the
treatment presented in the Section \ref{sectionII} and
\ref{SectionIII}, the local field effects of the induced dipoles at
the interfaces need to be properly treated. This separation of the
macroscopic and the microscopic anisotropy of the interfacial
monolayer is the ensurence of the simplicity and effectiveness in
interpretation of the surface SHG and SFG-VS data.

Of course, above discussion may not be fully suitable for the
situation when the interface consists of a big number of layers of
the anisotropic nonlinear induced dipoles. However, as long as such
film is thin enough not to significantly alter the macroscopic
reflection and transmission coefficients of light from the interface
between the two isotropic phases, there is no reason that one should
worry about the issue of the linear macroscopic dielectric constant
of the film itself.\cite{RoyPRB200013283,ShenNonlinearOpticsBook}

\subsection{The planewise dipole sum rule and the local field factor in the interface layer}

The treatment of the microscopic local field effect as expressed in
the Eq.\ref{Localfieldfactor} in the Section \ref{sectionII} is
similar to the results as Ye and Shen's earlier
work.\cite{YePRB19834288} The only difference is that now the
general contribution from layers other than the interface monolayer
is now included. To implement these formulae to correctly calculate
or estimate the local field factors, the planewise dipole sum rule
in the molecular crystal dielectric theory need to be
invoked.\cite{SchacherPR1965,
MahanPR1969,PhilpottJCP1972588,PhilpottJCP1972996,PhilpottJCP1973595,PhilpottJCP19745306}
The implementation of the planewise dipole sum rule in studying the
interface problem was discussed by Munn \textit{et al.} in the
1990s.\cite{MunnJCP199310052,MunnJCP199310059,PanhuisJCP20006763,PanhuisJCP200010685,PanhuisJCP200010691}
However, its implication has not been picked up by others in the SHG
and SFG-VS community so far. These issues needs to be addressed.

Now let us look at the actual problems in the interpretation of the
SHG measurement data. In their SHG studies of the
self-assembled-monolayer (SAM) at the gold substrate, Eisert
\textit{et al.} carefully compared the results with the macroscopic
three layer model and the two layer model, as well as the local
field corrections. They concluded that using the two-layer model
without local-field correction gave most satisfactory agreement of
the molecular orientation with the results from the NEXAFS and IR
spectroscopy measurements.\cite{EisertPRB199810860} According to
Eisert \textit{et al.}, the parameters used to calculate the local
field factors in the Eq.\ref{Localfieldfactor} and the
Eq.\ref{alphaexpression} of the pNA [O$_{2}$N-C$_{6}$H$_{4}$-]
monolayer in the SHG experiment are: $a=5$ to $5.5{\AA}$,
$\alpha_{aa}=\alpha_{bb}=16.4 {\AA}^{3}$, $\alpha_{cc}=27.3
{\AA}^{3}$, and $\theta \sim 52^{\circ}$ from NEXAFS measurement.
Then, one has $\alpha_{xx}=\alpha_{yy}=19.8 {\AA}^{3}$ and
$\alpha_{zz}=20.5 {\AA}^{3}$, then $l_{xx}=l_{yy}=3.51$ to $2.16$
and $l_{zz}=0.403$ to 0.473. Therefore, the effective microscopic
dielectric constant defined as $\varepsilon'=n'^{2}=l_{yy}/l_{zz}$
becomes 8.71 to 4.57, or $n'=2.95$ to 2.14. Here the definition
$\varepsilon'=n'^{2}=l_{xx}/l_{zz}$ follows the definition by Wei
\textit{et al.}.\cite{WeiXing2000PRE62p5160} These are unreasonably
large values and the calculation by Eisert {et al.} concluded that
in order to get the $\theta \sim 52^{\circ}$ from the SHG data, $n'$
has to be close to unity, i.e. neglecting the microscopic local
field effect.

Such conclusion by Eisert \textit{et al.} is in direct disagreement
with the formulation provided by Shen \textit{et
al.}\cite{WeiXing2000PRE62p5160} However, the formulation with the
microscopic local field factors by Shen \textit{et al.} is validated
by the treatment in this work. The answer to Eisert \textit{et
al.}'s difficulties in calculating the local field factors lies in
the planewise dipole sum rule,\cite{SchacherPR1965,
MahanPR1969,PhilpottJCP1972588,PhilpottJCP1972996,PhilpottJCP1973595,PhilpottJCP19745306}
which Munn \textit{et al.} used to tackle the problem of
overestimation of the local field factors by putting the linear
polarizability of the whole molecule into the
Eq.\ref{Localfieldfactor}.\cite{MunnJCP199310052,MunnJCP199310059,PanhuisJCP20006763,
PanhuisJCP200010685,PanhuisJCP200010691}

The planewise dipole sum rule was originally developed in explaining
optical and dielectric phenomena in molecular crystals, such as the
long range coupling of the exciton in the molecular crystal, in the
early 1970s.\cite{SchacherPR1965,
MahanPR1969,PhilpottJCP1972588,PhilpottJCP1972996,PhilpottJCP1973595,PhilpottJCP19745306}
It has been known that the plane sum in the summation of the
dipole-dipole interaction as in the Eq.\ref{electric filed of
dipole}, which leads to expressions of the local field factors in
the Eq.\ref{Localfieldfactor}, falls off exponentially as the
perpendicular distance from the plane of the origin
increases.\cite{PhilpottJCP1972588,PhilpottJCP1973595} Philpott
\textit{et al.} showed that even for the strong dipoles in the
molecular crystals, the contribution of the immediate neighboring
layer is only less than $1\%$ of the contribution of the plane of
the origin, and the total sum of all the rest of the layers up to
infinity distance other than the plan of origin is generally less
than $10\%$.\cite{PhilpottJCP1972588,PhilpottJCP1973595}

Based on this planewise dipole sum rule, when the lateral distance
between the molecules is smaller than the length of the molecule,
there is no reason to view the molecule as a whole when calculating
the local field factors using the Eq.\ref{Localfieldfactor}. Munn
\textit{et al.} thus proposed a bead model to divide the long chain
into a chain of sphere beads in calculating the local field factor
for each submolecular segment layer for the long chain molecules in
the Langmuir-Blodgett
film.\cite{MunnJCP199310052,MunnJCP199310059,PanhuisJCP20006763,
PanhuisJCP200010685,PanhuisJCP200010691} This bead model is also
unknowingly supported by the calculations by Ye and Shen on the
contribution of the local field factors from the image induced
dipoles at a vacuum/metal interface.\cite{YePRB19834288} They
conclude that for $\varepsilon_{metal}=10$, i.e.
$(\varepsilon-1)/(\varepsilon+1) \sim 1$, when $d/a > 1/2$, the
image induced dipole contribution is negligible. Here $d/a > 1/2$ is
just the sphere bead condition in Munn's treatment if the
interfacial induced dipole and its image induced dipole are
considered as a whole unit.

In our treatment, the planewise dipole sum rule is explicitly
described with the Eq.\ref{dipolesumoftheirlayerconvergent}.
Evaluation of the $\xi_{0}^{l}$ term indicates that in most of the
cases, even the immediate neighboring term, i.e. $l=\pm 1$, is often
negligible when $d$ is larger than $3{\AA}$. This indicates that
when dealing with the local field factor calculations with the
Eq.\ref{Localfieldfactor}, the depth of the segmentation of the
molecular layer is generally not more than $3{\AA}$. However, this
number may not be correct for the big chromophores with conjugate
structures, which may require quantum mechanical treatment or more
complex models rather than the simple classical dipole model.

With the planewise dipole sum rule in mind, now one can understand
why in the Eisert \textit{et al.} case\cite{EisertPRB199810860} that
the local field factors are over estimated than the actual value.
Because the SAM monolayer is closely packed, if one use the
$\alpha_{ii}$ value of the whole pNA group, the $l_{ii}$ factors
should be more deviated from the unity value. If one use the sphere
bead model, the $l_{ii}$ is going to be much more close to unity.
Actually, if closely inspect the formulae in the
Eq.\ref{Localfieldfactor}, one can expect that when
$\mid\alpha_{ii}\xi_{0}/2a^{3}\mid > 1$, the $l_{xx}$ and $l_{yy}$
can even become a large negative value. If the sphere bead model is
not used, a closely packed long chain molecular monolayer would
easily have unreasonable local field factors according to the
Eq.\ref{Localfieldfactor} without considering the planewise sum
rule. Therefore, the planewise dipole sum rule is very important and
has to be implemented for the evaluation of the microscopic local
files factors in the molecular monolayer.

One direct conclusion from the segmentation of the interfacial
molecules for the implementation of this planewise dipole sum rule
is that because the different groups in the same molecule has
different linear polarizabilities, and because the same molecular
group may have different polarizabilities at different frequencies,
their local field factors can be significantly different when these
groups occupies different segmented layers when the monolayer is
closely packed. This immediately poses questions to the practice of
using the same and simple value for the interface local field
factors in the SHG and SFG-VS
studies.\cite{ZhuangPRB199912632,SimpsonAnalChem2005215,FreyCPL2000454}
One of the example is for the SHG and SFG-VS studies of the 5CT
[CH$_{3}$-(CH$_{2}$)$_{4}$-(C$_{6}$H$_{4}$)$_{3}$-CN] monolayer at
the air/water interface.\cite{ZhuangPRB199912632} In the
interpretation of the SHG measurement of the
-(C$_{6}$H$_{4}$)$_{3}$-CN chromophore and the SFG-VS measurement of
the -CH$_{3}$ and -CN groups of the 5CT monolayer, the same local
field factor value, i.e. $n'=1.18$, was used for all three molecular
chromophore or groups. Close inspection shows that the orientational
angle thus obtained is actually inconsistent with other
studies.\cite{BiadaszLC20041639,HertmanowskiJMS2005201,ZDSWHF5CBand5CT,ZhenZJPCB200112118}
Therefore, whether the same $n'=1.18$ value is suitable for the
three groups needs to be re-examined. In a recent study in our
research group, using a self-consistent approach we found that the
$n'(800nm)\sim 1.5$ and $n'(400nm)\sim 2.5$ for the
-(C$_{6}$H$_{4}$)$_{3}$-CN chromophore from the SHG data,
significantly different from that of the -CH$_{3}$ or -CN group, and
consistent with the large polarizability of the
-(C$_{6}$H$_{4}$)$_{3}$-CN
chromophore.\cite{ZDSWHF5CBand5CT,ZDSPhDDissertation} The
orientational angle thus obtained for the -(C$_{6}$H$_{4}$)$_{3}$-CN
chromophore in the closely packed Langmuir monolayer at the
air/water interface is about $20^{\circ}$, consistent with the
orientational angle obtained in other
studies.\cite{BiadaszLC20041639,HertmanowskiJMS2005201}

Now, if the $\varepsilon'$ or $n'$ values should be different, why
in the past SFG-VS studies it was quite successful to use the simple
values such as $n'=1.18$, or the average value of the optical
constant of the two neighboring bulk phases for molecular groups and
chromophores at the common dielectric
interfaces?\cite{WangHFIRPC2005,WangHFPCCP2006,
ZhuangPRB199912632,SimpsonAnalChem2005215,FreyCPL2000454,ZhenZJPCB200112118}
Here we can show that these values are actually quite reasonable
approximation of the common simple molecular groups such as
-CH$_{3}$, -C=O, etc. at the interface.

The experimental linear polarizability values of many common
molecular group are as listed in the Table
\ref{PolarizabilityValues}.\cite{MillerJACS19797206,MillerJACS19908533,MillerJACS19908543}
The values for simple organic groups are typically in the range of
1.5 to 3.0 ${\AA}^{3}$. Considering that the segmentation depth is
expected to be not more than $3{\AA}$ for simple molecules with no
big chromophores with conjugated structure, as a realistic
approximation, the typical $\alpha$ value used in the calculation is
set in between 2 to 4 ${\AA}^{3}$. Considering the fact that at the
molecular interface a typical molecular group per area is about 20
${\AA}^{2}$ (square lattice constant $a \sim 4.5 {\AA}$) to 30
${\AA}^{2}$ ($a \sim 5.5 {\AA}$), simulations in Table
\ref{simulationLocalfieldfactors} shown that the typical
$\varepsilon'$ value for these groups are 1.17-1.56, i.e.
$n'$=1.08-1.32, depending on the lattice constant and the value of
the polarizability.

In above calculation we assumed that the monolayer is closely
packed. If the interfacial monolayer is sparse, the local field
effect shall be small. For the situation of submonolayer adsorption
at the liquid-liquid interface, then the dipole interactions of the
solvent molecules in the interface plane need to be considered.

These values are generally in between the vacuum and the substrate
dielectric constant, as used in the literature. They are also close
to the estimation made by Zhuang \textit{et al.} using a modified
Lorentz model of the interface,\cite{ZhuangPRB199912632} and the
simple average of the optical constant of the two adjacent isotropic
bulk
phases.\cite{SimpsonAnalChem2005215,FreyCPL2000454,RowlenAnalChem20025954}
However, even though these values worked almost fine, the models and
the reasonings behind it prevented the treatment of the molecular
details of the molecular interface. With the development of the
laser techniques and the measurement methodology in the surface SHG
and
SFG-VS,\cite{WangHFPCCP2006,WangHFIRPC2005,GanweiJPCC2007p8716,GanweiJPCC2007p8726}
now such details can be interrogated from the careful experiments.
This shall provide new opportunities for the interface studies.

\begin{table}[h!]
\caption{Experimental polarizabilities values for common molecular
groups from the literatures.\cite{MillerJACS19908533}}
\begin{center}
\small{
\begin{tabular}{lcccccccccccccc}
\hline
  ~group~ & ~$\alpha({\AA}^{3})$~ & ~group~ & ~$\alpha({\AA}^{3})$~\\
\hline
  ~-COOH~ & ~2.86~ &  ~-SH & ~3.47~ \\
  ~-CH$_{3}$~ & ~2.24~ &  ~-CN~ & ~2.16~ \\
  ~-CH$_{2}$-~ & ~1.84~ &  ~-NH$_{2}$~ & ~1.76~ \\
  ~-C=O~ & ~1.82~ &  ~H$_{2}$O~ & ~1.45~ \\
\hline\label{PolarizabilityValues}
\end{tabular}
}
\end{center}
\end{table}

\begin{table}[h!]
\caption{Simulation of local field factors $\varepsilon'$ and $n'$
with the $\alpha$ value between 2 and 4 ${\AA}^{3}$. The case for
$\alpha=1.5 {\AA}^{3}$ is the simulation results of the interfacial
water molecules.}
\begin{center}
\small{
\begin{tabular}{lcccccccccccccc}
\hline
  ~$\alpha({\AA}^{3})~$ & ~$a({\AA})$~ & ~$l_{xx}=l_{yy}$~ & ~$l_{zz}$~ & ~$\varepsilon'$~ & ~$n'$~ \\
\hline
  2.0 & 4.5 & 1.11 & 0.84 & 1.33 & 1.15\\
  2.0 & 5.5 & 1.06 & 0.90 & 1.17 & 1.08\\
  4.0 & 4.5 & 1.25 & 0.72 & 1.74 & 1.32\\
  4.0 & 5.5 & 1.12 & 0.82 & 1.37 & 1.17\\
  1.5 & 3.5 & 1.19 & 0.76 & 1.56 & 1.25\\
  1.5 & 4.0 & 1.12 & 0.83 & 1.36 & 1.16\\
\hline\label{simulationLocalfieldfactors}
\end{tabular}
}
\end{center}
\end{table}

Of course the above simple estimation does not include the bigger
chromophore, such as phenyl, and biphenyl etc. Their polarizability
is generally much larger, and the local field effects can be much
stronger when they are closely aligned at the interface. According
to the Eq.\ref{Localfieldfactor}, it is possible to have the
$l_{xx}=l_{yy}$ to be smaller than unity or even go to negative
values, and $l_{zz}$ to be larger than unity, when the molecules
with relatively larger linear polarizability are closely packed even
far from resonance. These are the unique phenomena for the two
dimensional anisotropic local field factors. According to the
Lorentz relation, this is not possible for the optical bulk
materials with normal dispersion. Such phenomena are certainly yet
to be explored experimentally.

The local field factors in the molecular layer is dependent on the
orientation and also on how the segmentation is done, and one
generally does not have the detailed knowledge of these information
beforehand. Therefore, different approximation approaches were
employed in the past SHG and SFG-VS studies. Similar to what Munn
\textit{et al.} did
previously,\cite{MunnJCP199310052,MunnJCP199310059,PanhuisJCP20006763,
PanhuisJCP200010685,PanhuisJCP200010691} we invoked the planewise
dipole sum rule here to show that in calculating the local filed
factors using the microscopic point dipole model using formulae in
the Eq.\ref{Localfieldfactor}, extreme care must be taken in order
not to overestimate the local field effects. However, this does not
imply that the sphere bead model always generates more reasonable
results for the local field factors. We want to point out that when
the square lattice constant is close to or even smaller than the
size of the molecule under studying, the simple point dipole model
is expect to breakdown, and the planewise dipole sum rule can be
used as the remedy to some extent. One expects that there is no
simple rule to segment the chain molecules in calculating the local
field factors. Down to the detailed molecular level, quantum
mechanic treatment of the electron density and their polarizability
should be employed to make more accurate description of the
molecules at the interface. Nevertheless, the above estimation ca
give the reasonable upper and lower bound of the local field
factors.

\subsection{SHG or SFG-VS experiments to test the microscopic model}

Here we discuss how SHG and SFG-VS experiment can be used to test
the microscopic model.

The key point raised in this study is that the molecular optics
treatment of the interface nonlinear radiation can provide a
detailed microscopic description of the nonlinear optical processes
and the molecular behaviors at the interface. The key to test the
microscopic model is to find way to determine the microscopic local
field factors or to quantitatively determine their effects in the
SHG or SFG-VS experiment, and to compare these experimental effects
with the molecular details in the molecular monolayer.

There were few attempts trying to experimentally determine the
interfacial effective optical constants with
SHG\cite{SimpsonAnalChem2005215,FreyCPL2000454} and
SFG-VS.\cite{ChenJCPB2004} All of them concluded for a value close
to the simple arithmetic average of the optical constant of the two
adjacent bulk phases. These works were not based on the microscopic
model and they can not be used to test the microscopic model.
Because the microscopic model predicts that the local field factors,
which is connected to the effective optical constant of the
interface layer through the relationship
$\varepsilon'=n'^{2}=l_{yy}/l_{zz}$,\cite{ZhuangPRB199912632,WeiXing2000PRE62p5160}
must depend on the detailed structure of the molecular monolayer and
may have values other than the simple arithmetic average or the
prediction with the modified macroscopic Lorentz model of the
interface.\cite{ZhuangPRB199912632}

The first test is to determine the local field factor
$l_{yy}/l_{zz}$ for the closely packed Langmuir monolayers at the
air/water interface with large chromophores. According to the
Eq.\ref{Localfieldfactor}, when $\alpha_{ii}\sim 15 {\AA}^{3}$ and
$a^{2}\sim 40 {\AA}^{2}$, one has that
$\varepsilon'=l_{yy}/l_{zz}\sim 2.10$, thus $n'\sim1.45$. These kind
of conditions can be easily satisfied with chromophore with two or
three phenyl groups, such as the alkyl cyano-biphenyl (nCB) or the
alkyl cyano-triphenyl (nCT) Langmuir films. It has been shown that
in the closely packed 8CB Langmuir monolayer at the air/water
interface, since the molecular surface density is always known, a
self-consistent analysis of the polarization dependent SHG data at
different surface density can give the molecular orientation angle
and the $\varepsilon'$ value at different surface
densities.\cite{WangRaoJCP2003,ZDSWHF5CBand5CT} Such self-consistent
analysis was suggested by Munn \textit{et al.} before, because they
realized that the determination of the molecular orientational angle
and the determination of the interface local field factor are
interconnected.\cite{PanhuisJCP200010691} Such self-consistent
analysis can only be carried out for the Langmuir or
Langmuir-Blodgett monolayers because their surface density are known
and can be varied controllably. It turned out in our SHG studies
that for the 8CB monolayer, the $\varepsilon'$ changes from 1.7 to
2.4 when the surface density changes from 51${\AA}^{2}$ to
39${\AA}^{2}$ per molecule.\cite{ZDSWHF5CBand5CT,ZDSPhDDissertation}
These results agree very well with the Eq.\ref{Localfieldfactor},
and the orientational angle thus obtained agree well with the rod
model of the chromophore. This is a very convincing example to show
that the local field factors of the molecular monolayer do not have
simple values close to the arithmetic average of the two adjacent
bulk phases. The detail of this study is to be published
elsewhere.\cite{ZDSWHF5CBand5CT,ZDSPhDDissertation}

Another test can be designed with the SFG-VS measurement of the
small rigid linear molecules, such as the acetonitrile molecule, at
the air/liquid interface. For example, SFG-VS can selectively
measure polarization dependent vibrational spectra of both the
$-CH_{3}$ and the $-CN$ groups of the interface acetonitrile
molecule.\cite{Eisenthal1993JCP5099} Since the molecule is linear,
the orientation angles, which are dependent on the values of
$\varepsilon'_{-CN}$ and $\varepsilon'_{-CH_{3}}$ respectively,
determined from the $-CH_{3}$ and the $-CN$ SFG-VS data should give
the same value. Thus the ratio
$\varepsilon'_{-CN}/\varepsilon'_{-CH_{3}}$ can be determined
experimentally. Using the polarization null angle method in
SFG-VS,\cite{WangHFIRPC2005,GanWeiCPLSensitivity,LurongCSB2003,Chenhua2005CPL}
the ratio of the local field factor for the $-CH_{3}$ and the $-CN$
groups, i.e. $\varepsilon'_{-CN}/\varepsilon'_{-CH_{3}}$, can be
obtained quite accurately. This value is indeed different from unity
and it is beyond the experimental error bar. This observation can
not be explained by the previous models, but can be well explained
and quantitatively analyzed with the microscopic model in this work.
The detail of the experimental study and analysis is going to be
reported elsewhere.\cite{ZhangZhenToBePublished}

More SHG or SFG-VS experiments can be designed accordingly to test
the microscopic model. Detail analysis of SHG and SFG-VS data is now
possible with the recent development of quantitative polarization
measurement in SHG and
SFG-VS.\cite{WangHFIRPC2005,WangHFPCCP2006,Simpson2005PRB125409,Simpson2003ACA133}
These studies not only can provide test for the microscopic model,
but also can provide detailed information of the molecular interface
according to the microscopic model.

\section{Conclusion}

Because the SHG and the SFG-VS have been proven as the sensitive
probes for interfaces with the submonolayer
coverage,\cite{ShenNature1989,ShenARPC1989,Corn1994ChemRev,Eisenthal1996ChemRev}
the treatment based on the more realistic discrete induced dipole
lattice model is developed in this report. Recent development in
quantitative polarization and symmetry analysis in SHG and SFG-VS
have provided new opportunities in the interface
studies.\cite{WangHFIRPC2005,WangHFPCCP2006} These development also
provided more accurate data to test the SHG and SFG-VS theories at
the detailed molecular
level.\cite{GanweiJPCC2007p8716,GanweiJPCC2007p8726,WangRaoJCP2003,SimpsonAnalChem2005215,ZhuangPRB199912632}

In this report we have taken the molecular optics theory approach
and treated the molecular interface with more realistic discrete
induced dipole lattice model. Based on this model, the following
results are derived: a. the detailed expressions of the local field
factors in the interface molecular layer; b. the detailed
expressions for the far field radiation of the SHG as well as the
SFG-VS process from the interface induced dipoles. It turned out
that the asymptotic results for the far field radiation is in the
same form as the results derived from the famous infinitesimally
thin polarization sheet layer by Heinz and Shen using the Maxwell
equation with the boundary
conditions,\cite{Heinzthesis1982,ShenARPC1989} as well as their
later modifications with the consideration of the microscopic local
field factors using the classical dipole
model.\cite{ZhuangPRB199912632,WeiXing2000PRE62p5160,YePRB19834288}
This not only validates the effectiveness of the original
formulation from the works by Shen and colleagues, but also
validates the success of the molecular optics theory approach
developed here.

According to Born and Wolf, the molecular optics theory can directly
connect the macroscopic optical phenomena to the molecular
properties, and can provide deeper physical insight into
electromagnetic interaction problems than does the rather formal
approach based on Maxwell's phenomenological
equations.\cite{Bookmaxborn1997,WolfJOSA1972} Based on the
microscopic and discrete induced dipole lattice model, the problem
of the macroscopic dielectric constants of the interfacial molecular
monolayer is discussed and clarified. We explicitly demonstrated
that the macroscopic dielectric constant of the molecular
submonolayer and monolayer does not need to be invoked in the
microscopic model of the surface nonlinear optics. Based on the
planewise dipole sum rule and the previous work by Munn \textit{et
al.}, the issue on how the microscopic local field factors are
evaluated was discussed. According to these results, the
effectiveness and limit of the simple models on the effective
dielectric constant of the interfacial molecular monolayer are
discussed and evaluated. However, for simple and small molecular
groups, the simple models does provide good approximation of the
microscopic local field effects of the interfacial molecular
monolayer. SHG and SFG-VS experimental tests of the microscopic
discrete induced dipole lattice model are also discussed. With these
developments, many previous SHG and SFG-VS studies are better
understood and evaluated. Moreover, these developments can provide a
full microscopic description of the nonlinear radiation from the
molecular interface. Detailed molecular information can be obtained
from the model developed in this work, together with better and more
accurate polarization measurement data in SHG and SFG-VS.

The microscopic molecular optics approach can also be applied to the
treatment of the ellipsometry response from the molecular monolayer
interface. When ellipsometry is to be applied to study such
interfaces, a microscopic molecular theory for the ellipsometry is
certainly needed. So far, the treatment on the ellipsometry and the
treatment of the SHG or SFG-VS are not fully consistent in
description of the anisotropy in the molecular layer. We shall
discuss these issues elsewhere.

In conclusion, aside from the macroscopic theory of the surface SHG
and SFG-VS, an effective microscopic molecular optics theory is
developed in this work. Such development may shed new light on both
the linear and the nonlinear optics interface studies.

\vspace{0.5cm}

\noindent \textbf{Acknowledgment.} DSZ thanks the helpful discussion
from Wen-kai Zhang, Yu-jie Sun and Yuan Guo. HFW thanks the support
by the Natural Science Foundation of China (NSFC, No.20373076,
No.20425309, No.20533070).

\renewcommand\baselinestretch{2}

\end{document}